\documentclass[iop]{emulateapj}

\usepackage{amsmath}
\usepackage{subfigure}
\usepackage{graphicx}
\usepackage{natbib}
\usepackage{hyperref} 
\usepackage{float}

\graphicspath{{figuresall/}}

\newcommand{\bi}{\begin{itemize}}
\newcommand{\ei}{\end{itemize}}
\newcommand{\beq}{\begin{equation}}
\newcommand{\eeq}{\end{equation}}
\newcommand{\bea}{\begin{eqnarray}}
\newcommand{\eea}{\end{eqnarray}}
\newcommand{\bqu}{\begin{quote}}
\newcommand{\equ}{\end{quote}}
\newcommand{\bctr}{\begin{center}}
\newcommand{\ectr}{\end{center}}
\newcommand{\bd}{\begin{description}}
\newcommand{\ed}{\end{description}}
\newcommand{\bdm}{\begin{displaymath}}
\newcommand{\edm}{\end{displaymath}}

\newcommand{\om}{\Omega_m}

\newcommand{\hinvMpc}{$h^{-1}$Mpc}
\newcommand{\hinvGpc}{$h^{-1}$Gpc}

\newcommand{\OM}{\Omega_{\rm M}}
\newcommand{\OL}{\Omega_{\Lambda}}

\newcommand{\OMout}{\Omega_{\rm M}^{\rm out}} 
\newcommand{\OLout}{\Omega_{\Lambda}^{\rm out}} 
\newcommand{\OMin}{\Omega_{\rm M}^{\rm in}} 
\newcommand{\lcdm}{$\Lambda$CDM }

\slugcomment{}

\shorttitle{Low-$z$ cutoff in supernova cosmology}
\shortauthors{Sinclair, Davis, Haugb{\o}lle}

\begin{document}

\title{Residual Hubble-bubble effects on supernova cosmology}

\author{
Benjamin Sinclair\altaffilmark{1}, 
Tamara M.\ Davis\altaffilmark{1,2},
Troels Haugb{\o}lle\altaffilmark{3}
}

\altaffiltext{1}{Dept. of Physics, University of Queensland, QLD 4072, Australia}
\altaffiltext{2}{Dark Cosmology Centre, Niels Bohr Institute, University of Copenhagen, Denmark}
\altaffiltext{3}{Niels Bohr Institute, University of Copenhagen, Denmark}

\begin{abstract}
Even in a universe that is homogeneous on large scales, local density fluctuations can imprint a systematic signature on the cosmological inferences we make from distant sources. One example is the effect of a local under-density on supernova cosmology.  Also known as a Hubble-bubble, it has been suggested that a large enough under-density could account for the supernova magnitude-redshift relation without the need for dark energy or acceleration.  Although the size and depth of under-density required for such an extreme result is extremely unlikely to be a random fluctuation in an on-average homogeneous universe, even a small under-density can leave residual effects on our cosmological inferences.  It is these small under-densities we consider here.

In this paper we show that there remain systematic shifts in our cosmological parameter measurements, even after excluding local supernovae that are likely to be within any small Hubble-bubble.  We study theoretically the low-redshift cutoff typically imposed by supernova cosmology analyses, and show that a low-redshift cut of $z_0\sim0.02$ may be too low based on the observed inhomogeneity in our local universe.  

Neglecting to impose any low-redshift cutoff can have a significant effect on the cosmological parameters derived from supernova data.  A slight local under-density, just 30\% under-dense with scale $70$\hinvMpc, causes an error in the inferred cosmological constant density $\OL$ of $\sim4$\%.  Imposing a low-redshift cutoff reduces this systematic error {\em but does not remove it entirely}.  A residual systematic shift of 0.99\% remains in the inferred value $\OL$  even when neglecting all data within the currently preferred low-redshift cutoff of 0.02.  Given current measurement uncertainties this shift is not negligible, and will need to be accounted for when future measurements yield higher precision.
\end{abstract}

\keywords{cosmology: observations, theory --- supernovae: general}



\section{Introduction}\label{sect:introduction}
To estimate global properties of the universe, such as the average matter density, we make models that include those properties as parameters and test which parameter values best fit the data. The simplest cosmological model that provides a good fit to supernova data is the \lcdm model, in which the universe is assumed to be homogeneous and dominated by cold dark matter and a cosmological constant, parametrized by their normalized densities $\OM$ and $\OL$.  The assumption of homogeneity however, is only valid over large scales. Evidently, on the scales of stars, galaxies, and clusters of galaxies the matter distribution is far from homogeneous. Such inhomogeneities give rise to varying expansion rates in different parts of the universe, and varying gravitational potentials. 

Thus, as light makes its journey from source to observer it experiences not a cosmological redshift due to a single rate of expansion $\bar{H}(t)$, but rather varying degrees of cosmological redshift determined by the local rate of expansion $H(r,t)$. In addition, the spatially varying gravitational potential contributes differing gravitational red- and blue-shifts, as the light travels through under- and over-dense regions. 

To a large extent these effects cancel each other out over the path of the photon.  In a flat universe containing only matter the canceling is exact, however in the presence of other contributions, such as from dark energy or curvature, the potential wells evolve as the light passes through them in such a way so as to alter the total redshift.  This effect is known as the Integrated Sachs-Wolfe (ISW) effect and has been recently used to detect the presence of dark energy from temperature enhancements in the CMB correlated with the density of galaxies along the line of sight \citep{giannantonio08}.  

Nevertheless the ISW effect is small (an order of magnitude below the level of the primordial fluctuations in the CMB) and as we look over a range of directions we benefit from spatial averaging which further smooths any variation due to density fluctuations along any particular line of sight.  Spatial averaging also compensates for  the over- or under-density at the sources.

The one irreducible effect comes from any density fluctuation inside which we happen to reside.  Then the absolute depth of the potential (rather than just the change in depth during a light-crossing time) comes into full effect.  In that case the photon has no chance to climb back out of the potential well and feels the full brunt of the density fluctuation.  
Whether we be in an under-dense or over-dense region, it is clear that the fluctuation light travels through immediately before reaching us is not compensated for, as it would have been were the light left to continue its journey.

This final redshift will add a systematic shift in the magnitude-redshift data to which we fit our homogeneous models, and may impede our ability to deduce the average density of the universe outside our local bubble.

We are therefore motivated to study the possible effect of a local inhomogeneity on our cosmological inferences.  In this paper we focus on supernova cosmology.  We generate simulated data with a local under-density then fit it with an homogeneous cosmological model to test whether the derived parameters match the input parameters for the average density outside the local void (Sect.~\ref{sect:results}).  Supernova cosmologists typically reject low-redshift data on the basis that peculiar velocities are relatively high compared to the cosmological recession velocities and Hubble bubble effects most prominent.  We therefore implement a low-redshift cutoff, denoted $z_0$, and investigate how well the fit converges to the input model as this low-redshift cutoff changes (Sect.~\ref{sect:lowz}).  

Initially we use a large void in a matter-only universe to demonstrate the concept clearly, then in Sect.~\ref{sect:largeLambda} introduce a cosmological constant as well.  In Sect.~\ref{sect:quantifying} we assess the likely impact of voids of the size we expect to find in random fluctuations in a standard $\Lambda$CDM cosmology.  Our conclusions are analyzed in Sect.~\ref{sect:conclusion}.  We now begin by first discussing the nature and sizes of voids in our universe.

\section{Void sizes}\label{sect:void}

The word `void' has been used in the literature to mean several different things.  The `voids' we refer to here are gaussian under-densities, of the kind you might expect to naturally occur as random fluctuations in a generally homogeneous universe.  While very large voids are non-gaussian in nature, as a first approximation, and for small voids, the assumption of gaussianity should hold.  Our motivation is to study the possible effect of a local density fluctuation of a size that is likely to occur in a standard $\Lambda$CDM cosmology.  We choose to concentrate on under-densities rather than over-densities because of the interest they have already generated in the community: firstly due to the possibility of mimicking a cosmological constant (for very large scale under-densities) and secondly because of the observational indications from supernova cosmology and galaxy surveys that there may be a local void (a `Hubble bubble').

The evidence for a local Hubble bubble was first found by \citet{zehavi98}, in which the supernova data appeared to show that $H_0$ within 70\hinvMpc\, was 6.5\%$\pm$2.2\% higher than the value outside that distance (presuming a flat $\OM=1.0$ model).  They explained this by a 20\% under-density surrounded by a dense shell, and noted that this size corresponds roughly to the size observed for local large scale structure  \citep{geller97}, such as the `great walls'.  \citet*{jha07} refined this observation with more supernovae, and considered a void embedded in a $\Lambda$CDM universe.   They find the significance of the \citet{zehavi98} result drops to $\delta_H =(H_{\rm in}-H_{\rm out})/H_{\rm out} \sim $4.5\%$\pm$2.1\% in the $\Lambda$CDM case and find a similar value of 6.5\%$\pm$1.8\% with their new data.  

These results have since been challenged by several papers, such as \citet{conley07} who showed that using a different light-curve-fitter with a different color treatment can remove the evidence of a void.  Data from other distance-measurement techniques do not find a significant Hubble bubble.  Tully-Fisher measurements by \citet{giovanelli99} find a statistically insignificant Hubble bubble of $\delta_H\sim$1\%$\pm$2\% while \citet{hudson04} use the peculiar velocity field within 120\hinvMpc\, to find modest evidence of locally enhanced Hubble expansion with $\delta_H\sim 3\pm1.3$\%.

These observations have all been measurements of the local expansion rate.  Alternatively one can look to large galaxy redshift surveys and measure directly from the observed structure the typical size of voids in the galaxy distribution.  \citet{hoyle04} found using the 2dFGRS that 35\% of the universe consists of voids larger than 10\hinvMpc, with a mean galaxy under-density of $\bar{\delta}_{\rm gal}=(\rho_{\rm in}-\rho_{\rm out})/\rho_{\rm out}=-0.94\pm0.02$. 
Fluctuations in the dark matter density will be less extreme than this, since this is a measurement of the {\em galaxy} density fluctuation, which is enhanced due to galaxy bias (galaxies are at the denser parts of the dark matter distribution).  

Perhaps the most relevant observations are those of our local universe.  In a survey of our local universe, out to a redshift of $z\sim0.15$, \citet{geller97}  find evidence for inhomogeneity on the 100\hinvMpc\, scale.  Earlier \citet{geller89} noted that voids of 50\hinvMpc\, with density only $\sim 20$\% of the mean are ubiquitous to all surveys, and the largest local structure, the `Great Wall' has dimensions at least 60\hinvMpc\, by 170\hinvMpc.

Theoretical calculations of the void size that is likely in a $\Lambda$CDM universe \citep[e.g.][]{furlanetto06} find void sizes and density contrasts somewhat smaller than these and than the 2dFGRS survey observations.  This remains a point of tension.  Moreover, recent observations of bulk flows in the universe, such as those by \citet{kashlinsky08} using the Sunyaev-Zeldovich effect in the CMB, show that large-scale motions also appear to be larger than predicted by theoretical $\Lambda$CDM calculations and simulations.  

This tension between observation and theory, married with the hint of a Hubble bubble in the supernova data and the observed distance to the `great walls' of our local structure, suggests that a void as large as 70\hinvMpc\, is a reasonable, and interesting, void size to test.  The wide range of density contrasts considered in the literature makes it difficult to define a `typical' under-density.  The depth of under-density depends strongly on the size of void considered. We choose a value of $\delta=-0.3$ (corresponding to $\rho_{\rm in} = 0.7\rho_{\rm out}$) for all the simulated voids in this paper. The typical peculiar velocity corresponding to such a void is of the order of 100km s$^{-1}$. Taking this as a monopole velocity at a scale of 70\hinvMpc\ it is a typical (a $\sim$1-1.5 sigma level) fluctuation in a standard $\Lambda$CDM model \citep[][]{haugbolle07}, and having the additional coincidence of being near the Centre of the void with correlated monopole velocities out to at least 70 \hinvMpc\ does not lower this probability much more than to the 2 sigma level.
 
The precise value of the size and density contrast of the void we test is not of prime importance.  We focus on the qualitative results which remain roughly the same regardless of choice of void size.  In particular we concentrate on how a void of any size affects supernova cosmology and whether we can remove its effect by neglecting data within the void. 

An alternative cosmological model has recently emerged \citep[see e.g.][and references therein]{enqvist08}, where we are at the center at a gigaparsec sized void embedded in an Einstein-de Sitter universe. It has been shown that while we have to give up the Copernican Principle, and fine tuning is needed to have the observer very near the center of the void, such a model is viable when compared to current state-of-the-art cosmological observations 
\citep{alnes06a,garcia-bellido08,garcia-bellido08-kz,garcia-bellido09,zibin08,clifton08,alexander09}.  That includes not only supernova, but also CMB and BAO observations \citep{garcia-bellido09}, and a void of such dimensions, would be difficult to spot in galaxy redshift surveys, due to the small gradient in the density profile. The model is attractive because there is no dark energy and the observed late-time acceleration is a consequence of a larger Hubble rate near the center of the local under-density. The main drawback is the required coincidence of having the Milky Way near the center to make the CMB radiation from the surface of last scattering appear close to isotropic \citep{alnes06b}.

\begin{figure}
\epsscale{1.0}
\plotone{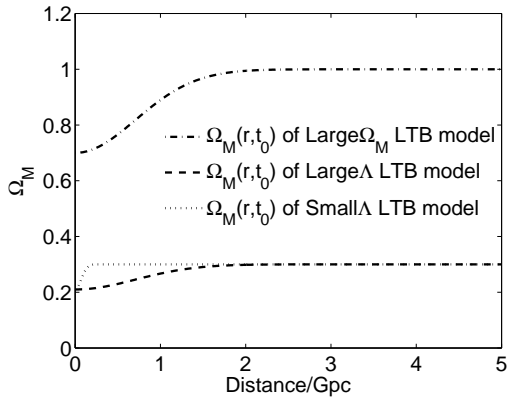}
\caption{A snapshot at $t_0$ of the radial profile of the matter density $\OM(r,t_0)$ used to generate the simulated data sets in Sects.~\ref{sect:largeVoid}-~\ref{sect:quantifying}.}
\label{fig:gaussianLTB}
\end{figure}

 We begin by considering a large void of scale size $r_0=700$\hinvMpc\, to demonstrate the concept clearly.  A void of this order of magnitude is what would be required to explain the supernova data without a cosmological constant \citep{alexander09}.  We embed this large void in both a flat $\om=1.0$ universe (without a cosmological constant, Sect.~\ref{sect:largeVoid}), and in a $\Lambda$CDM universe with $(\OM,\OL)=(0.3,0.7)$ (Sect.~\ref{sect:largeLambda}).  
We assess the likely impact of a more realistic sized void of $r_0=70$\hinvMpc\, in Sect.~\ref{sect:quantifying}.   A summary of the void sizes we consider appears in Table~\ref{table:voids}, and is depicted in figure~\ref{fig:gaussianLTB}.

\begin{deluxetable}{lccccc}
\tablecolumns{6}
\tablecaption{Void models used.}
\tablehead{
\colhead{} & \colhead{$\OMout$} & \colhead{$\OL^{\rm out}$} & \colhead{$h_{\rm out}$} & \colhead{$r_0$ (Mpc/$h$)} & \colhead{$\delta\,^{\dag}$ }
}
\startdata
Large $\Omega_M$ & 1.0 & 0.0 & 0.7 & 700 & -0.3 \\
Large $\Lambda$ & 0.3 & 0.7 & 0.7 & 700 & -0.3 \\
Small $\Lambda$ & 0.3 & 0.7 & 0.7 & 70 & -0.3 \\
\enddata
\tablenotetext{0}{$^\dag$$\delta\equiv(\OMin-\OMout)/\OMout$}
\label{table:voids}
\end{deluxetable}

\section{Investigating the homogeneous universe in the presence of a local under-density}\label{sect:method}

In this section we explain the method we use to measure the effect of a local inhomogeneity on the cosmological parameters derived using supernova cosmology.  We generate simulated data containing an under-dense region at the origin, and then investigate the impact of neglecting to account for the under-density by fitting the data with an homogeneous \lcdm model.  The impact of the void is the difference between the best-fit cosmological parameters derived using the homogeneous model and those input into the LTB simulation.

\subsection{Modeling a local under-density}
The model we use to describe a local void is a Lema\^{i}tre-Tolman-Bondi
(LTB) model, which is isotropic but inhomogeneous. The metric of our
model can be written
\begin{equation}\label{eq:metric}
ds^2 = -dt^2 + \frac{(a' r + a)^2}{1 - k(r) r^2} dr^2 + a^2 r^2 d{\Omega}^2\,,
\end{equation}
where $a(r,t)$ is an effective scale factor, $k(r)$ describes the curvature
as a function of coordinate distance $r$, and prime denotes differentiation with respect to $r$.

Defining the transverse and longitudinal Hubble parameters
$H(r,t) \equiv\frac{\dot a}{a}$ and $H_L(r,t)\equiv\frac{\dot a' r + \dot a}{a' r + a}$
from the Einstein equations one can construct an effective Friedmann equation
\citep[see][]{garcia-bellido08}
\begin{align}\label{eq:friedmann}
H^2(r,t) &= H_0^2(r)\left[\OM(r) \left(\frac{a_0(r)}{a(r,t)}\right)^3 \right. \\ \nonumber
  & + \left. (1-\OM(r)-\OL(r))\left(\frac{a_0(r)}{a(r,t)} \right)^2  + \OL(r) \right]
\end{align}
where $H_0(r) = H(r,t_0)$, and $a_0(r)=a(r,t_0)=1$ is a gauge freedom
giving the scale factor at $t_0$. The total matter density $\OM(r)$ and the
cosmological constant $\OL(r)$ are related to $H_0(r)$, the curvature $k(r)$, and the
physical matter density $\rho_M$ at $t=t_0$ as,
\begin{align}
\OM(r) & = \frac{8 \pi G}{H_0^2(r) a_0^3(r) r^3} \int_0^r dr' r'^2 \rho_M(r',t_0), \\
\OL(r) & = \frac{8 \pi G}{3 H_0^2(r)} \rho_\Lambda, \\
k(r) & = H_0^2(r) (\OM(r) + \OL(r) - 1) a_0^2(r).
\end{align}
The time to big bang as a function of distance is a function of radius, which can
be seen by integrating (\ref{eq:friedmann}) with respect to $a(r,t)$, but only models with a
constant time to big bang give a well motivated growing void profile as a function
of time \citep{zibin08}, and below we only consider such models.
 
Using the scale factor we can determine the time of emission for light at a given
redshift by solving the redshift equation,
\begin{equation} \label{eq:redshiftequation}
\frac{dt}{d(\log(1+z))}=- \frac{1}{H_L(r(z),t(z))}.
\end{equation}
We can also determine the luminosity distance (and hence apparent magnitude) from
\begin{equation} \label{eq:lumdist}
d_L(z)=(1+z)^2 a[r(z),t(z)] r(z).
\end{equation}
The distance modulus is given by $\mu = 5 \log_{10}(d_L(z)) + 25$, with luminosity distance measured in Mpc.

In our model the matter distribution at large distances is a constant $\OMout$, but nearby the density decreases to $\OMin$ at the origin. Thus, the LTB model reproduces the important features we are investigating: an on-average homogeneous universe in which we reside within a local Hubble-bubble. The LTB model can accept any density profile for the local inhomogeneity.  We choose to implement a simple gaussian, according to,
\begin{eqnarray}
\OM(r) &=& \OMout+(\OMin-\OMout)e^{-r^2/r_0^2},\\
 &=& \OMout(1+\delta e^{-r^2/r_0^2})\,,
\label{eq:gaussianLTB}
\end{eqnarray} 
where $\delta = (\OMin-\OMout)/\OMout = (\rho_M^{\textrm in}-\rho_M^{\textrm out})/\rho_M^{\textrm out}$ is the density contrast of the void. Given the density profile the Hubble parameter now $H_0(r)$ is uniquely defined up to a proportionality, due to the requirement of a constant time to big-bang \citep[][]{garcia-bellido08}.
We consider models in which the homogeneous universe outside the void consists exclusively of matter ($\OMout=1.0$) as well as models in which the universe exterior to the void is homogeneous \lcdm, $(\OMout,\OL^{\rm out})=(0.3,0.7)$.

In the latter case the mass density distribution is still given by equation~\ref{eq:gaussianLTB}, and the asymptotic cosmological constant density is
\begin{equation}
\OL(r\to\infty) = 0.7\,.
\label{eq:LTBLambda}
\end{equation} 
Close to the center of the void, the Hubble expansion rate is higher, while the value of the cosmological constant, $\rho_\Lambda$, is by construction constant, giving lower values for $\OL(r)$.

\subsection{Fitting an homogeneous model in the presence of an under-density}\label{sect:fitting}
The redshift and distance modulus calculated from equations~\ref{eq:redshiftequation} and~\ref{eq:lumdist} are the output from the LTB model, which we treat as the data we would collect if we were indeed situated at the center of a Hubble-bubble in an otherwise homogeneous (over large scales) universe. Explicitly, we input a set of $z$ values, which we take to be the redshifts of a set of supernovae we are observing, numerically integrate along a null geodesic in the LTB spacetime to obtain the position and time of emission from the source at redshift $z$, and then use this position and time to calculate a distance modulus, from equations~\ref{eq:friedmann} and~\ref{eq:lumdist}. To solve the equations we use an extended version of the easyLTB program \citep{garcia-bellido08}, that can handle an arbitrary mixture of dark matter and cosmological constant in the void. 

For the sake of finding the best fit cosmological parameters, we do not add simulated observational errors to the `correct' distance modulus, the reason for which is explained at the end of this section. To perform the $\chi^2$ minimizing fit, we simply assign an uncertainty to the distance modulus of $\pm$0.2 mag, which is a typical uncertainty in the measurement of distance modulus for supernovae. 

We then fit a set of parametrized homogeneous models to the LTB data and find which \lcdm cosmology is the best fit (by minimizing $\chi^2$).  This process is entirely analogous to the methodology cosmologists use to measure cosmological parameters from real data. The homogeneous models are parametrized by the values of (OM,OL, w) and we marginalize over H0 as per the standard procedure in supernova cosmology \citep{kim04}.

If the local inhomogeneity had no effect on our deductions then we would find that the best fit set of $\OM$ and $\OL$ would match the $\OMout$ and $\OL^{\rm out}$ we had input into our LTB model.

When we come to consider goodness-of-fit of the best fit models in section~\ref{sect:IncorrectConclusions} it is necessary to generate a set of observational errors to obtain realistic values of goodness-of-fit.  We generate observational errors according to a normal distribution with standard deviation 0.2 mag and add them to the distance modulus.  We repeat the procedure for different sets of normally distributed errors and average the result to get an mean goodness of fit, mean best fit matter density, mean best fit cosmological constant, and mean best fit equation of state. Note, that the errors are normally distributed, and therefore the mean best fit values of each parameter should converge (with infinite sets of different normally distributed errors) to the value obtained when no normally distributed errors were added.  This is the reason why we only need to add the observational errors when we want to estimate the goodness-of-fit; and why in sections ~\ref{sect:largeVoid} to ~\ref{sect:quantifying} we can avoid running multiple fits by performing a single fit without simulated errors, but giving each data point a 0.2 mag observational uncertainty.

\subsection{Sensitivity of Result to Size of Data Set}\label{sect:SizeOfDataSet}

It is important to note that the measure we use for goodness-of-fit will go down rapidly for the homogeneous model as the number of data points is increased and/or the precision of each data point is improved.  This is because an incorrect model (the homogeneous model in this case)  can not mimic the correct model in the limit of infinite data (or vanishing uncertainty)\footnote{As long as the model does not include the correct model in some limit of its parameters.}. 

This also means that the best fit parameters change as the number of data points varies.
When a model is a good description of the data, the data points will be evenly distributed above and below the theoretical curve.  However, when the model does not reflect the data there will be regions in which all data points lie above (or below) the theoretical curve.  (See the low-redshift regions of Fig.~\ref{fig:bestFitToGaussianLTB} for an example.)  Increasing the number of data points therefore gives more weight to the regions in which the model is the worst fit, since these regions have the most significant impact on $\chi^2$.  The best fit theoretical curve becomes distorted towards the data in the poor-fitting regions, at the expense of moving further away from the data in regions where the theoretical curve was a better fit.

Thus, when the test model is not an accurate representation of the system, as in our case of fitting a \lcdm model to LTB data, the cosmological parameters we infer will change depending on how many data points we choose to use. 

We have chosen to use 301 data points spread out uniformly with redshift over the redshift range $0 < z < 1.7$, since that is approximately the number available to current supernova surveys \citep[e.g.][]{kessler09}.\footnote{We could also choose to weight the redshift distribution according to the number of supernovae currently observed at each redshift, but that is unnecessary given the general nature of the analysis here.} 
As more supernova data are collected, and our uncertainty in their measurements reduces, it will become increasingly apparent whether an inhomogeneous or homogeneous model gives the better fit.

\section{Results: Neglecting to account for a local under-density}\label{sect:results}

\subsection{Large void in matter-only universe}\label{sect:largeVoid}

We begin by setting the void size to an extremely large $r_0 = 700$\hinvMpc. This is far larger than the expected scale of inhomogeneities in the standard \lcdm model, but is approximately the minimum size required to explain the supernova data without invoking a cosmological constant.  We use it here to clearly demonstrate the effect an under-dense region can have on our cosmological inferences.  

\begin{figure}
\epsscale{1.0}
\plotone{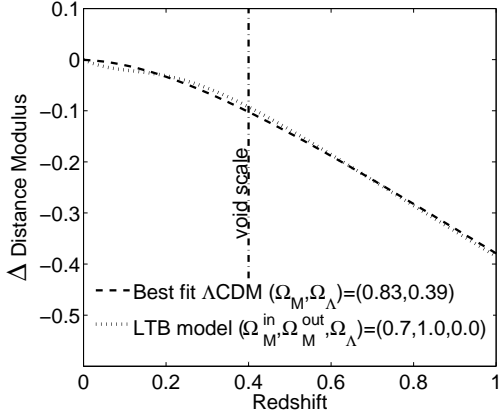}
\caption{In this Hubble diagram distance modulus relative to the empty universe is plotted against redshift.  This shows the \lcdm model (dashed line) that is the best fit to the magnitude-redshift data generated from the LTB model (dotted line) with $r_0=0.7$ \hinvGpc\ and $\OMin=0.7$ as used in Sect.~\ref{sect:largeVoid}.  The data was given an uncertainty of $0.2$mag.}
\label{fig:bestFitToGaussianLTB}
\end{figure}

We fit the standard \lcdm cosmology to our simulated void data and find the best fit to be
$(\OM,\OL)=(0.83\pm 0.08, 0.39)$. This model is plotted against the LTB data in Figure~\ref{fig:bestFitToGaussianLTB}.

The best fit matter density, $\OM=0.83$, is significantly lower than the actual value of $\OMout=1.0$ for the external universe in the LTB model. This simplified example demonstrates that a Hubble-bubble creates the illusion of a lower matter density than the true density outside the void.

The best fit model also has a significant cosmological constant, $\OL=0.39$, whereas the input model had none, $\OLout=0.0$. In section~\ref{sect:IncorrectConclusions}, we demonstrate that this extra parameter improves the goodness of fit significantly compared to a model with $\OL=0.0$. This demonstrates that the addition of an extra parameter (the cosmological constant) will be strongly supported when falsely attempting to fit an homogeneous model to data with such a large local void.\footnote{Note that we chose a void of this size specifically because of this feature.  To recover the $\Lambda$CDM parameters $(\OM,\OL)=(0.3,0.7)$ some fine tuning is required on the shape of the void.  The shape must be made sharper than our gaussian profile allows.}

\subsection{Large void in $\Lambda$CDM universe}\label{sect:largeLambda}

Having demonstrated the systematic shift that occurs if observations are made from a region of $\OM$ lower than the surroundings, we now show that the same behavior is expected in a universe with a cosmological constant.  

The model we now consider consists of an under-dense region embedded in the currently preferred \lcdm universe with $(\OM,\OL)=(0.3,0.7)$, which is a situation we may indeed find ourselves in if \lcdm is the correct description of the universe.  However, this model is still not `realistic' in the sense that we still use a large void size, $r_0=$700\hinvMpc, which is much larger than is likely to occur in a \lcdm universe, so the model is not self-consistent.  We use this model for comparison with the large void considered in the previous section.

\begin{figure}
\epsscale{1.0}
\plotone{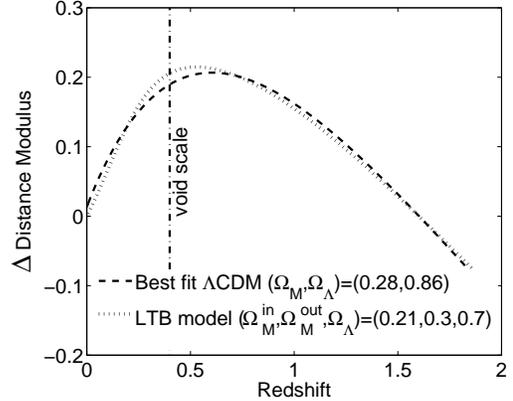}
\caption{This hubble diagram shows the \lcdm model (dashed line) that is the best fit to the magnitude-redshift data generated from the LTB model containing a cosmological constant(dotted line) with $r_0=0.7$ \hinvGpc\ and $(\OMin,\OMout)=(0.21,0.3)$ as used in Sect.~\ref{sect:largeLambda}.}
\label{fig:bestFitTobig_lambda}
\end{figure}

Figure~\ref{fig:bestFitTobig_lambda} depicts the best fit \lcdm model, which has $(\OM,\OL)=(0.28,0.86)$. This corresponds to a systematic error of 6.7\% and 23\% respectively.  
This example differs from the pure matter model tested in the previous section, because the homogeneous model we are fitting to the data has the same number of parameters as the simulated data exterior to the local void.  We find that in this case a non-zero $\Lambda$ is even more important to the fit than when there was no external $\Lambda$ in the simulation.  
In Table~\ref{table:AddingLambda} we show the results for the best fits of homogeneous models\footnote{Note, in Table~\ref{table:AddingLambda} we have used a different fitting methodology where we simulate observational errors, giving slightly different best fit parameters. The difference is within the estimated uncertainties of our best fit parameters.} with and without $\Lambda$.  In each case we show the $\chi^2$ values, indicating the goodness of fit.  In all cases the $\chi^2$ is much worse when no cosmological constant is included (as expected since adding parameters always allows a better fit), but the improvement in fit is more dramatic when the simulated universe included $\Lambda$.  That is as expected since in this case $\Lambda$ actually does relate to something real in the model.  In Sect.~\ref{sect:IncorrectConclusions} we study in more detail how much the additional parameters improve the fits, and whether the improvements are enough to fool us into believing false parameters.

\begin{deluxetable}{lcccccccc}
\tablecolumns{8}
\tablecaption{Improvements in $\chi^2$ by allowing $\Lambda$ and $w$ to vary}
\tablehead{
\colhead{Input model} & \colhead{Fitted model} & \colhead{$\OM$} & \colhead{$\OL$} & \colhead{$w$} & \colhead{$\chi^2$} & \colhead{dof} & \colhead{$\chi^2$/dof} & \colhead{AIC} }
\startdata
Large $\OM$	& CDM		& 0.87 & 0     & -1	 & 302.8  & 301-1 & 1.009 & 304.7  \\ 
Large $\OM$	& $\Lambda$CDM  & 0.85 & 0.41 & -1	 & 297.2  & 301-2 & 0.994  & 301.2\\ 
Large $\Lambda$ & CDM		& 0.18 & 0     & -1	 & 331.9  & 301-1 & 1.106  & 332.8\\
Large $\Lambda$ & $\Lambda$CDM  & 0.29 & 0.86 & -1	 & 297.6  & 301-2 & 0.995  & 301.5\\ 
Small $\Lambda$ & CDM		& 0.21 & 0     & -1	 & 320.1  & 301-1 & 1.067  & 322.0\\
Small $\Lambda$ & $\Lambda$CDM  & 0.31 & 0.72 & -1	 & 299.0  & 301-2 & 1.000  & 303.0 \\ 
Small $\Lambda$ & $w$CDM        & 0.25 & 0.75 & -1.7 & 298.2  & 301-3 & 1.001  & 306.2\\ 
Small $\Lambda$ & Flat $w$CDM   & 0.30 & 0.70 & -1.1 & 298.8  & 301-2 & 0.999  & 304.8   
\enddata
\label{table:AddingLambda}
\end{deluxetable}

The obvious question arises as to whether a local void can dupe us into believing in a dark energy equation of state that differs from the cosmological constant value of $w=-1$.  When we consider the more realistic sized voids in the next section we also fit a dark energy model in which $w$ is allowed to vary, $w$CDM.  This again restores us to the position of having one extra parameter that may allow a better fit.  In essence we are exchanging the $r_0$ parameter, which describes the size of the void in our input LTB model, with a false parameter, a free $w$.  The extra parameter is guaranteed to allow a better fit to the data.  The crucial point for us is whether the fit is improved so much that the extra parameter appears justified.  If so, then a local void is a significant danger for misleading our cosmological inferences.

\subsection{Small void in \lcdm universe}\label{sect:quantifying}
Having demonstrated that the Hubble-bubble alters the inferred external matter density and cosmological constant density, we now investigate the magnitude of this effect for a more modest sized under-density, of the kind that is predicted in self-consistent models of structure formation in \lcdm and seen in the typical size of structures actually observed in our universe.
The exact size of typical under-densities in our universe is the topic of ongoing debate, which we summarized in Sect.~\ref{sect:void}.  We have chosen to test a void with $r_0=$70\hinvMpc, and density contrast $\delta=-0.3$, as representative of the `likely' under-density we may find ourselves in. This corresponds to having a maximal monopole velocity of 120 km s$^{-1}$ inside the void, which is a typical monopole velocity for a shell in a standard \lcdm model \citep[][]{haugbolle07}.

Imprinting this void on a flat $(\OMout,\OLout)=(0.3,0.7)$ background, we find that the best fit \lcdm model has parameters ($\OM$,$\OL$)=($0.299$, $0.73$). 
The error in the best fit parameters is vastly reduced for the realistic sized under-density model compared to the large void model examined in Sect.~\ref{sect:largeLambda}, but is still large enough to be of concern. The errors in the best fit matter density  and best fit cosmological constant density are 0.3\% and 4.3\% respectively. These errors must be evaluated in the context of the current uncertainty in the cosmological parameters.  Using a combination of CMB, BAO and SN data \citet[][]{komatsu08} report $\OM h^2=0.1358^{+0.0037}_{-0.0036}$ and $\OL=0.726\pm0.015$ (Table 1, WMAP+BAO+SN mean values).  These correspond to errors of 2.7\% and 2\% respectively.  

Compared to the observational error in the cosmological constant density, the error incurred by fitting an inadequate model can be significantly larger. The corresponding systematic error in the matter density is small but not negligible compared to the current uncertainty, and as measurements become more precise the relative importance of this potential systematic error will increase.  In Sect.~\ref{sect:lowz} we discuss how this error is mitigated, but not removed, by introducing a low-$z$ cutoff on the data.

\subsubsection{Allowing $w\ne -1$}\label{sect:EoS}

One further cosmological inference on which we investigated the possible implications of  residing in an under-dense part of the universe, was the equation of state of dark energy. This is characterized by the parameter $w$, which relates the pressure and density of dark energy via $p=w{\rho}c^2$, such that $\rho \propto a^{-3(1+w)}$.

For dark energy that behaves as a cosmological constant, the value of $w$ is -1. This value is supported by supernova observations, for example \citep{WV07} use the ESSENCE supernova survey to constrain the value of $w$ to $1.07\pm0.09$  \citep[see also][]{astier06,kowalski08}. \citep[Although see][for recent developments showing a potential deviation from $w=-1$.]{kessler09,sollerman09} Here we examine whether a void could significantly change the inferred value of $w$ from the actual value. 

The LTB model we used as  a description of our universe was the same as that used in section~\ref{sect:quantifying}, with  $\OMin=0.21$, $\OMout=0.3$, $\OL=0.7$, and $r_0=70$\hinvMpc. In this LTB model the equation of state of dark energy is $w=-1$, so if upon fitting models with varying $w$, we find a best fit different to $w=-1$, we can say that this is solely due to the local under-density.

\begin{figure}
\epsscale{1.0}
\plotone{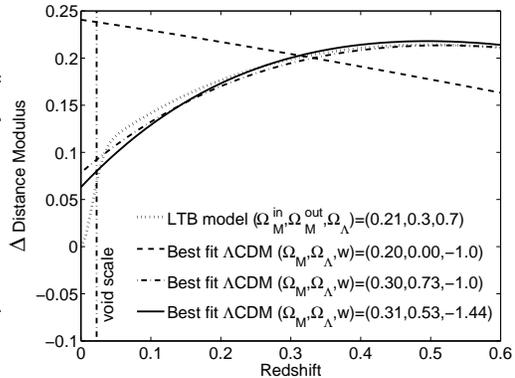}
\caption{In this figure we depict the effect of successively allowing extra parameters to vary. The dashed line shows the best fit model where only $\OM$ is allowed to vary, the dashed-dotted line where $(\OM,\OL)$ vary, and the solid line where all $(\OM,\OL,w)$ are free to vary.}
\label{fig:RLSuccessiveParameterAddition}
\end{figure}

At first, we simply add an extra parameter $w$, to our fitted models, so that each of $\OM$, $\OL$, and $w$ are allowed to vary freely and independently.  The error in $w$ is very large, 44\%, and the models fitted correspond very poorly to the external universe, the best fit parameters being $(\OM,\OL,w)=(0.314,0.53,-1.44)$. 
Such parameters are preferred since the low value of $w$ leads to a sharp drop in distance modulus as $z$ goes to zero. This mimics to some degree the effect of the void, depicted in figure~\ref{fig:RLSuccessiveParameterAddition}.

Since fitting for variable $w$ using only supernovae with no other constraints has large degeneracies, we now go on to demonstrate the behavior of the fitted value of $w$ under certain prior constraints.

First we adapt our fitting process to more closely correspond to conventional methodology. When determining the value of $w$, it is common practice to assume a flat universe. We introduce the constraint $\OM+\OL=1$, such that $\OL$ is no longer allowed to vary freely. With this restriction the error in $w$ is much reduced at only 7\%. The constraint on $\OM$ and $\OL$ means that low values of $w$ can no longer combine with the required $(\OM,\OL)$ to mimic the void at low redshifts, hence this undesirable behavior is avoided and the errors in $w$ are far smaller.

Thus we find that using a prior from other observations that constrains the $w$CDM model to be flat makes the result more robust to low-$z$ density fluctuations than fitting a general $w$CDM model with no prior.  However, one must also ensure that the prior used would not be rendered invalid by a putative local void.  

We note that the effect of a local void for the unconstrained and {\em flat} $w$CDM models is to push the best fit value of $w$ down, and so could give the illusion of phantom-like dark energy ($w<-1$) if the true value of $w=-1$.

Secondly we include a prior constraint on the value of $\OM=0.27{\pm}0.03$. This is included in our fits by modifying the $\chi^2$ estimate to $\chi^2_{\rm total}=\chi^2_{\rm SN}(\OM,\OL,w)+\chi^2_{\rm prior}(\OM)$,
where $\chi^2_{\rm prior}=\left(\frac{\OM-0.27}{0.03}\right)^2$. The best fit parameters under this constraint are $(\OM,\OL,w)=(0.27,0.86,-0.85)$. 
 Unlike the two previous varying-$w$ fitted models, the inferred $w$ increases and no longer suggests a phantom like dark energy.

\section{The low-redshift cutoff}\label{sect:lowz}
When fitting cosmological models to supernova data, it is common practice to remove data points of supernovae below a certain redshift. One reason is that at low redshifts the peculiar velocities of galaxies are a significant fraction of their recession velocities.  Peculiar velocities therefore add a large amount of scatter about the magnitude-redshift relation at low redshifts.  Typically supernova studies have neglected data from sources with a redshift below $z_0\sim0.02$ \citep{WV07,astier06,kessler09}, since at this redshift the mean peculiar velocities contribute less than 5\% scatter about the mean recession velocity.  

A potentially more significant reason from the perspective of the discussion in Section.~\ref{sect:introduction}, is that light from nearby sources does not give a good indication of the average density of the homogeneous universe, since much of its path has been through the local Hubble-bubble, should one exist.  This is potentially more significant than random peculiar velocities because it introduces a {\em systematic} shift onto all redshifts, not just a scatter.  The value of redshift below which we neglect data is referred to as the low-$z$ cutoff and we denote it by $z_0$. Figure~17 of \citet*{jha07} shows the derived value for the equation of state of dark energy, $w$, can change by as much as 20\% when the low-redshift cutoff is changed from $z_0\sim 0.008$ to $z_0 \sim 0.025$ in the presence of a void that extends to the larger of those two redshifts. 

A commonly held notion is that by neglecting data from sources that could be situated within a Hubble-bubble, we remove the impact of a local void \citep*[e.g.][footnote 20]{jha07}.  By neglecting nearby data the remaining light will have originated in, and primarily traveled through, the external region and thus should give a good indication of the external density.  
Here we investigate whether this low-$z$ cutoff is sufficient to satisfactorily mitigate the distorting effect of the local Hubble-bubble.
We want to know whether neglecting enough low-$z$ data allows us to recover the correct matter-density for the homogeneous region outside the void in our simulated data.  

We put this notion to the test by introducing a low-$z$ cutoff, and gradually increasing it to see if the derived cosmological parameters converge to the input parameters outside the void. 

We note from the outset that even the light originating from far away has to traverse the under-dense region before being observed, so removing nearby sources will not completely remove the effect of a void.

\begin{figure*}
\epsscale{0.45}
\begin{center}
\subfigure[Low-$z$ cutoff = 0.06]{\label{fig:varyingZ0-a}\plotone{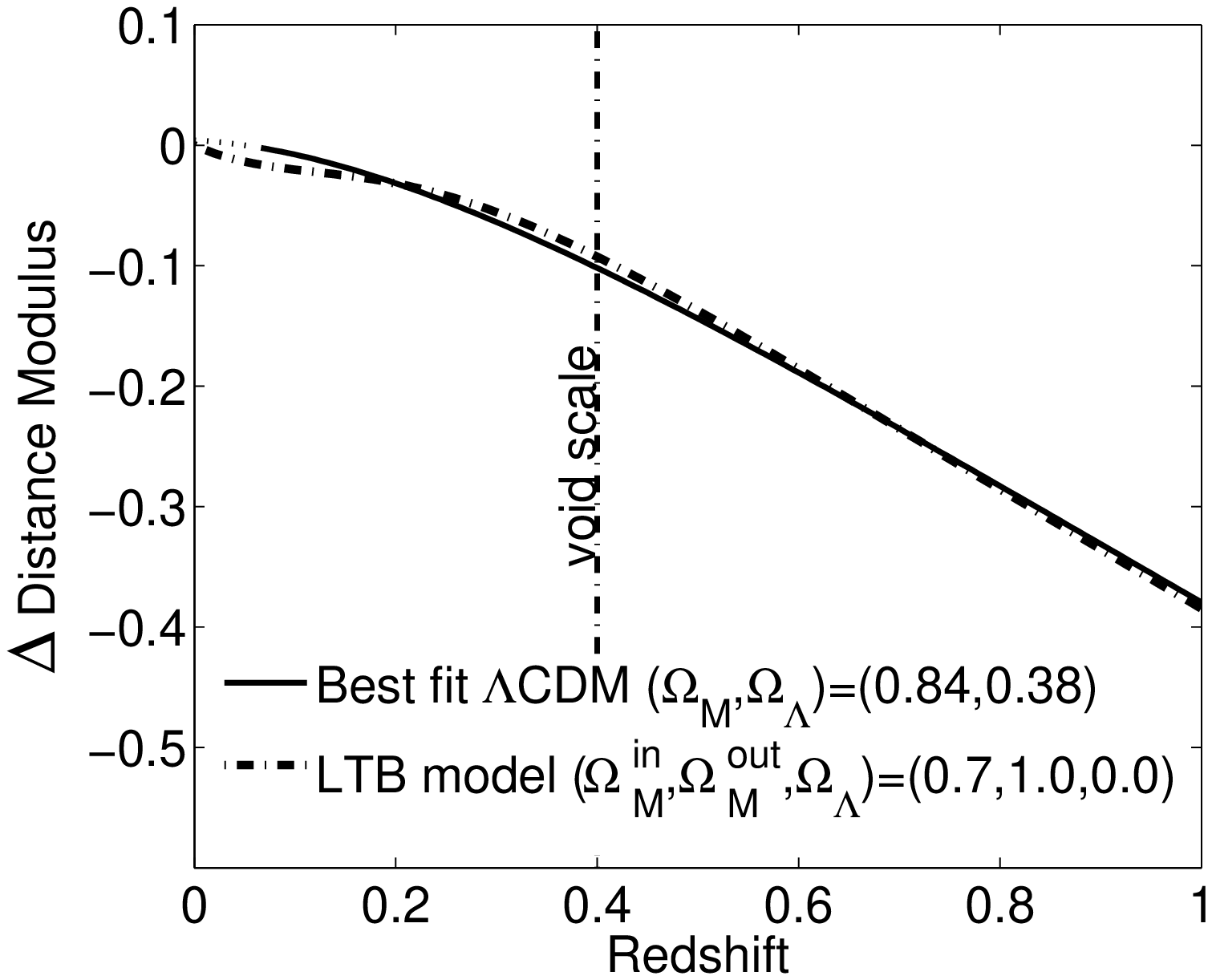}}
\epsscale{0.45}
\subfigure[Low-$z$ cutoff = 0.105]{\label{fig:varyingZ0-b}\plotone{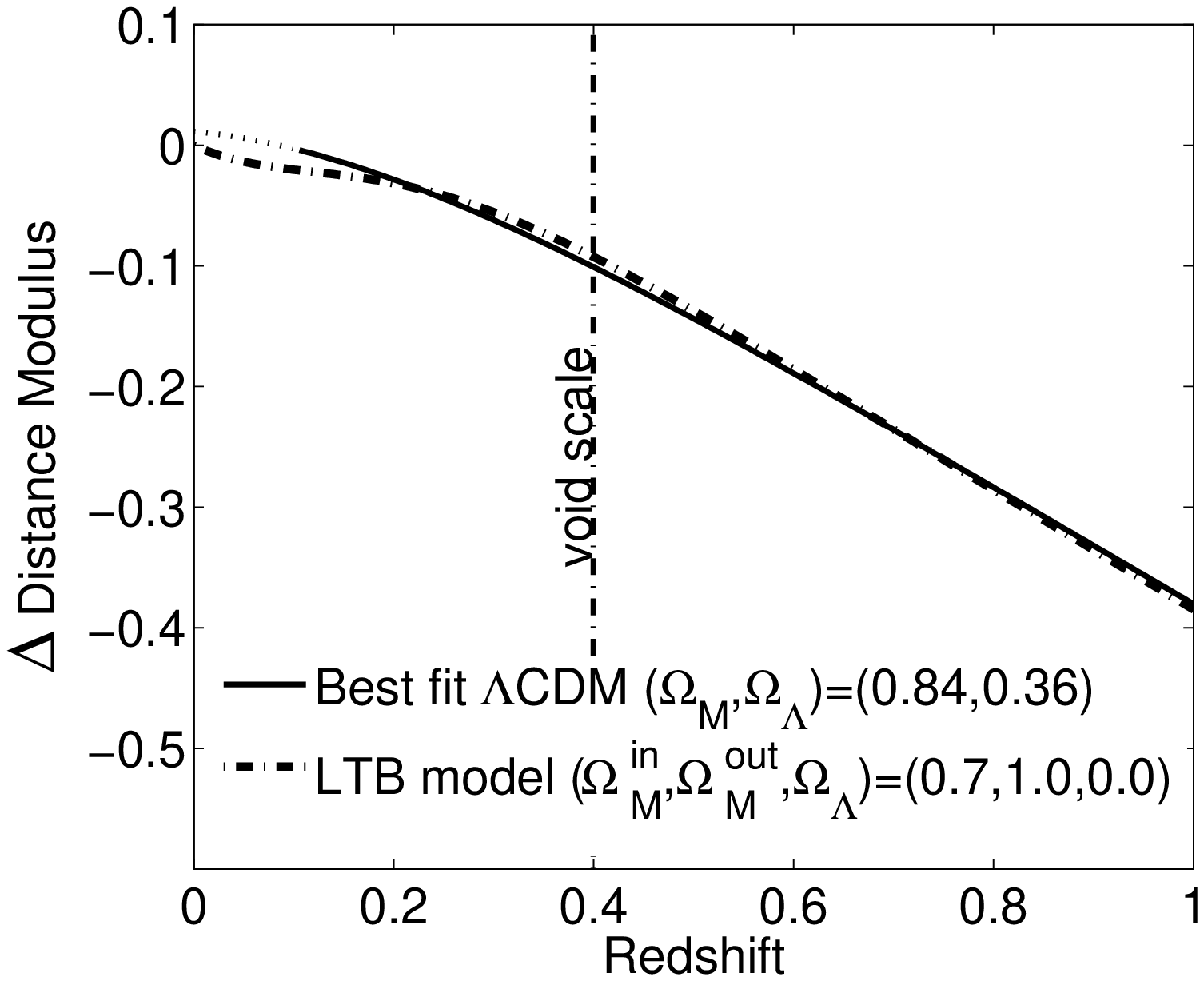}}
\epsscale{0.45}
\subfigure[Low-$z$ cutoff = 0.195]{\label{fig:varyingZ0-c}\plotone{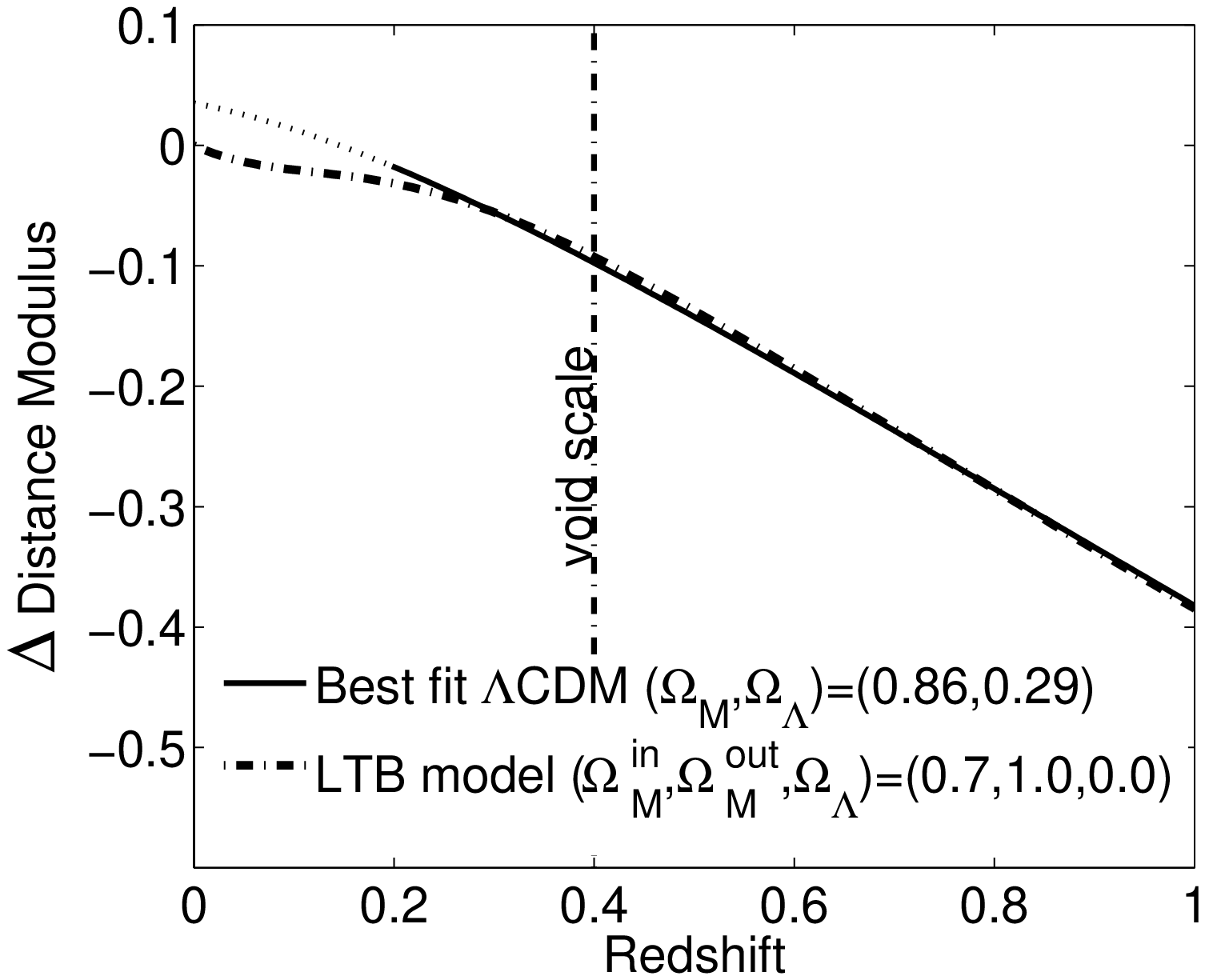}}
\epsscale{0.45}
\subfigure[Low-$z$ cutoff = 0.4]{\label{fig:varyingZ0-d}\plotone{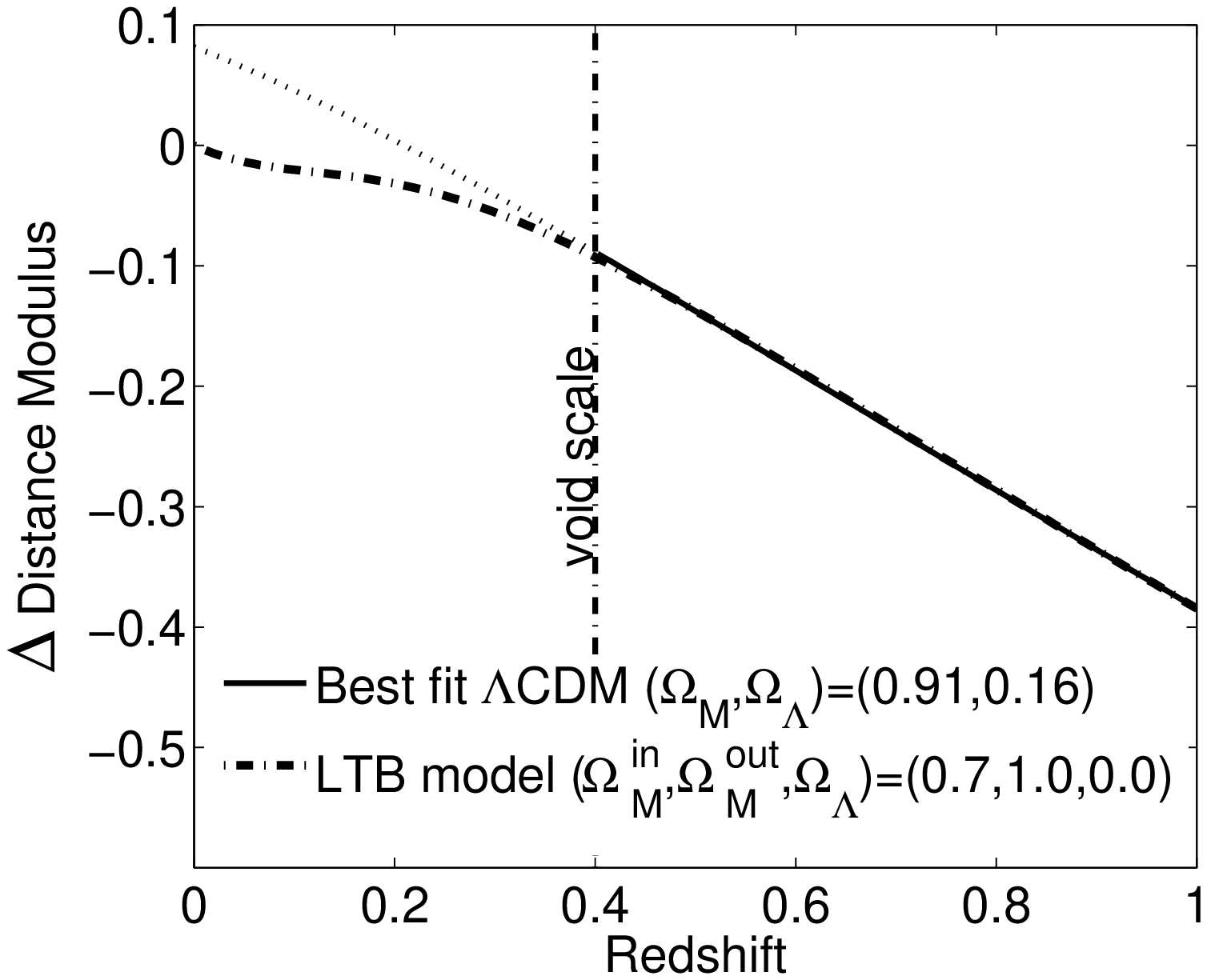}}
\end{center}
\caption{These plots of distance modulus (normalized to the empty universe) show how the best fit model changes as the low redshift cutoff changes.  Four models are plotted.  In each case they are plotted as a solid line over the range of redshifts used in the fit, and then extrapolated as dotted lines over the range of redshifts that were excluded from the fits.  As we implement progressively stronger low-$z$ cutoffs it is clear how the best fit model deviates ever more strongly from the data in the nearby under-dense region. }
\label{fig:varyingZ0}
\end{figure*}

Figure~\ref{fig:varyingZ0} shows four examples of different models that best fit the data as we move to progressively higher low-$z$ cutoffs, $z_0=[0.06,0.105,0.195,0.4]$ (Large-${\Omega}_M$ case).  It is clear that, as expected, the parameters of the best fit \lcdm models are in better agreement with the asymptotic values of the LTB model as we progressively neglect more low-$z$ data.  What may not be expected is that the best fit cosmological parameters do not converge to the known external density even after all data from within the Hubble-bubble have been rejected. 

\begin{figure}
\epsscale{1.0}
\plotone{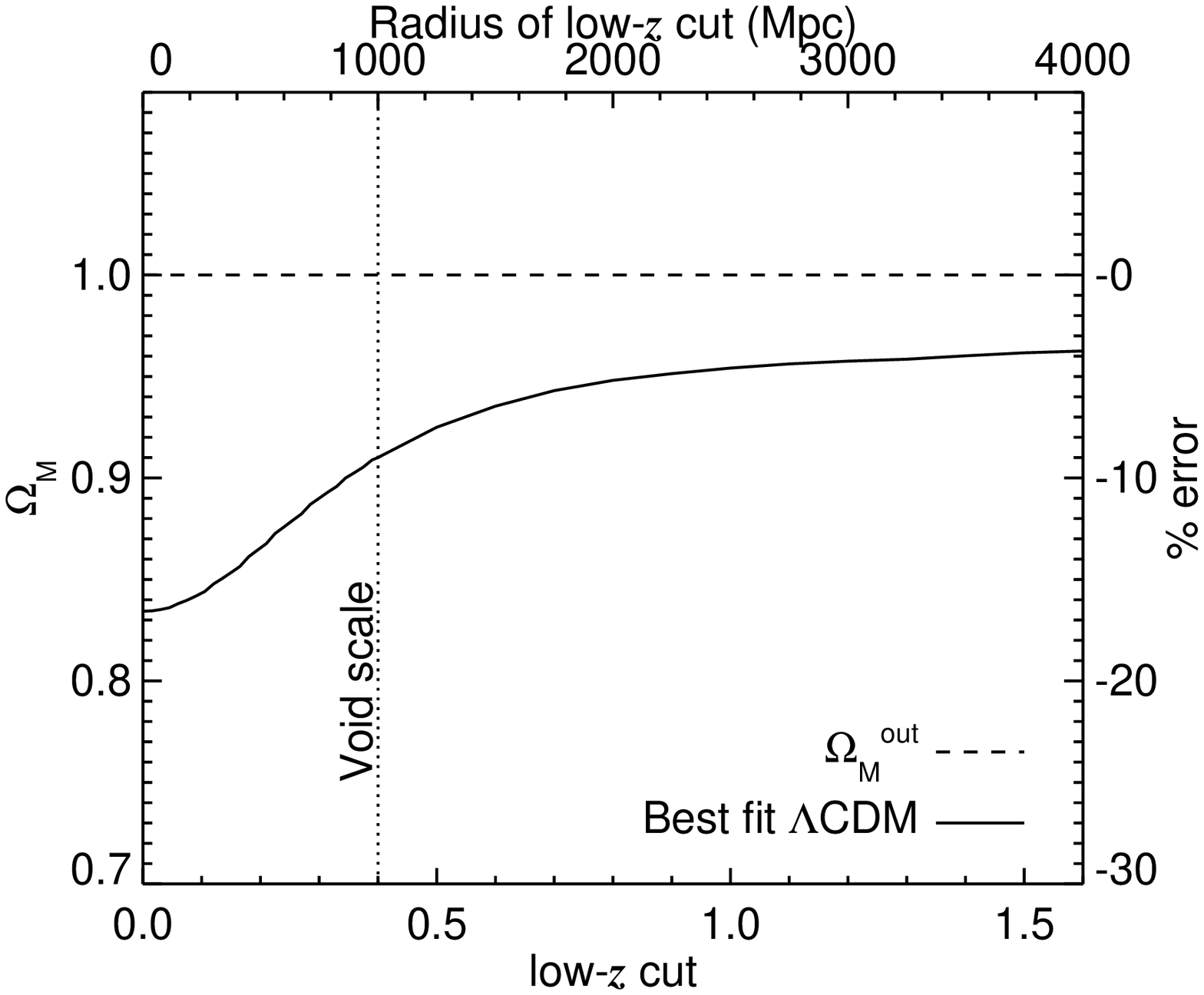}
\epsscale{1.0}
\plotone{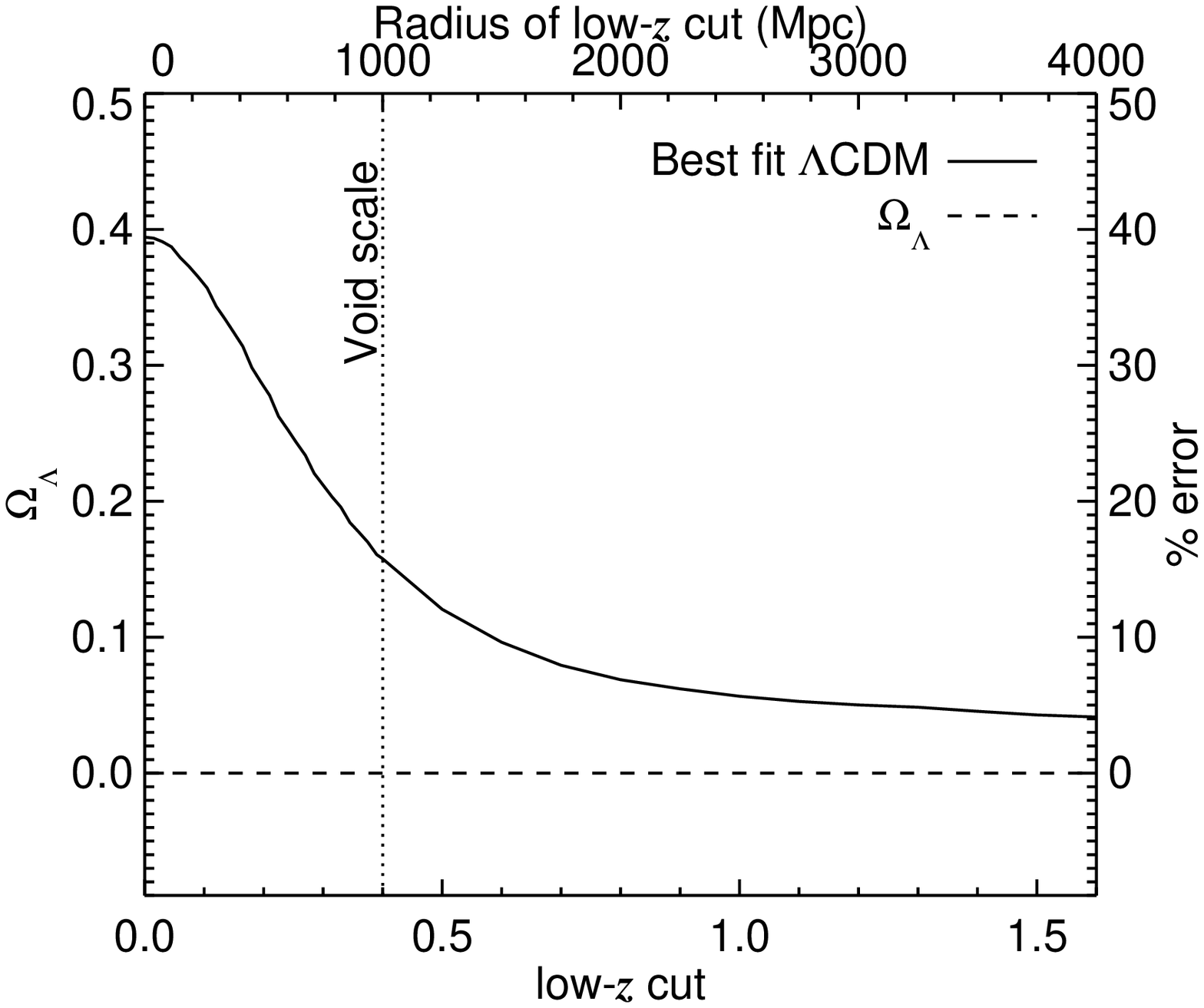}
\caption{Upper: As we progressively remove data points from low-$z$, the estimation of the external matter density improves, but does not converge to the true value.  Lower: \lcdm models with cosmological constant give a good fit to LTB models, despite there being no cosmological constant in the LTB model. However, as we remove low-redshift data points, this spurious cosmological constant is reduced.}
\label{fig:BigVoid_z0_d03}
\end{figure}

We summarize this in Figure~\ref{fig:BigVoid_z0_d03} 
which shows how the best fit $\OM$ changes with $z_0$. We see that as we remove data from sources within the Hubble-bubble, we do indeed get a better indication of the external matter density (which from our model we know to be $\OMout=1.0$). However, this figure also shows that the best fit $\OM$ asymptotes to a value less than $\OMout$.  Regardless of how much data we remove we cannot deduce the true external density in this manner.  

\begin{figure}
\begin{center}
\subfigure[Variation of $\Omega_{MBF}$ for a typical sized void]{\label{fig:RealLambda_OMz0_d03}\plotone{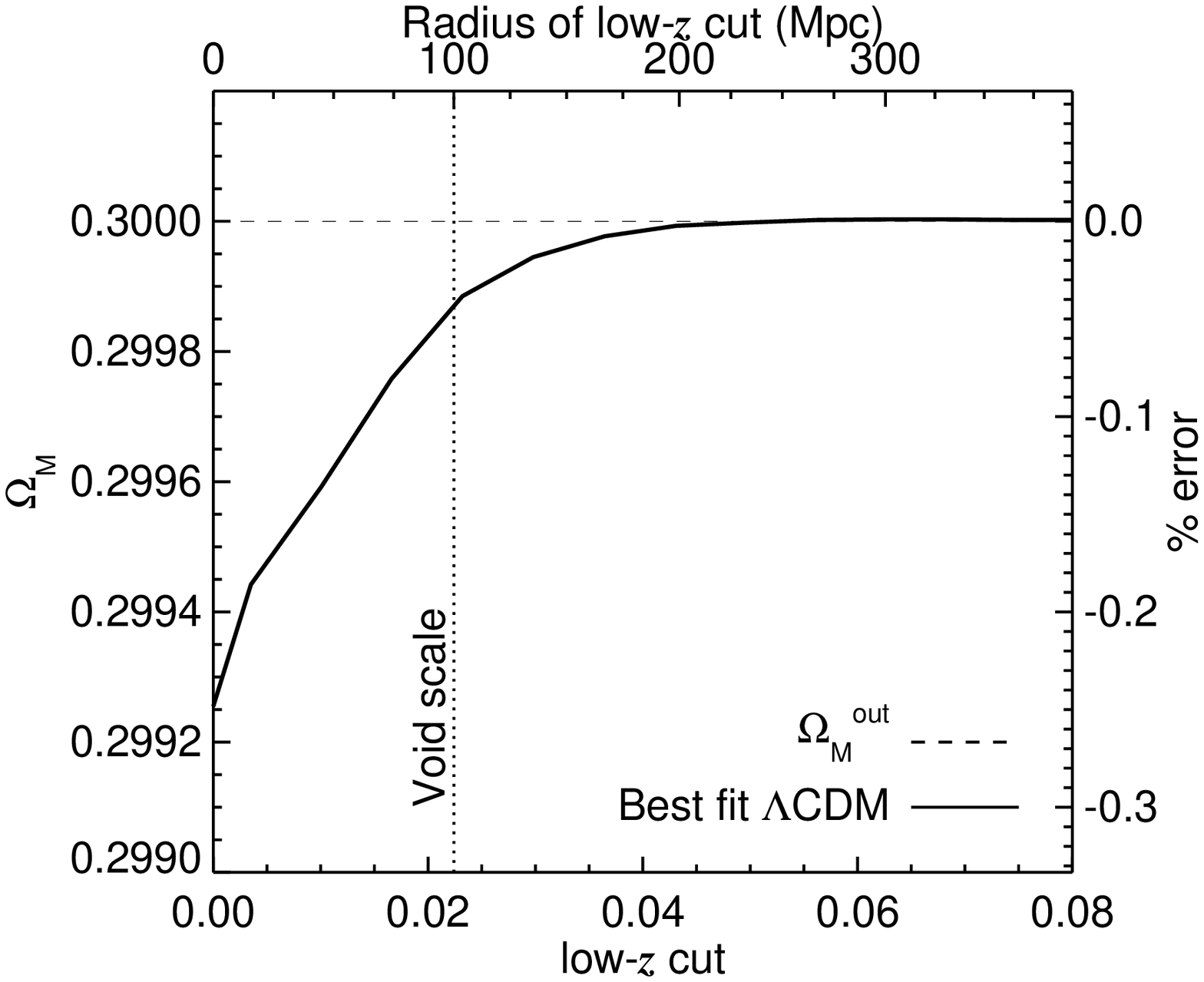}}
\subfigure[Variation of $\Omega_{{\Lambda}BF}$ for a typical sized void]{\label{fig:RealLambda_OLz0_d03}\plotone{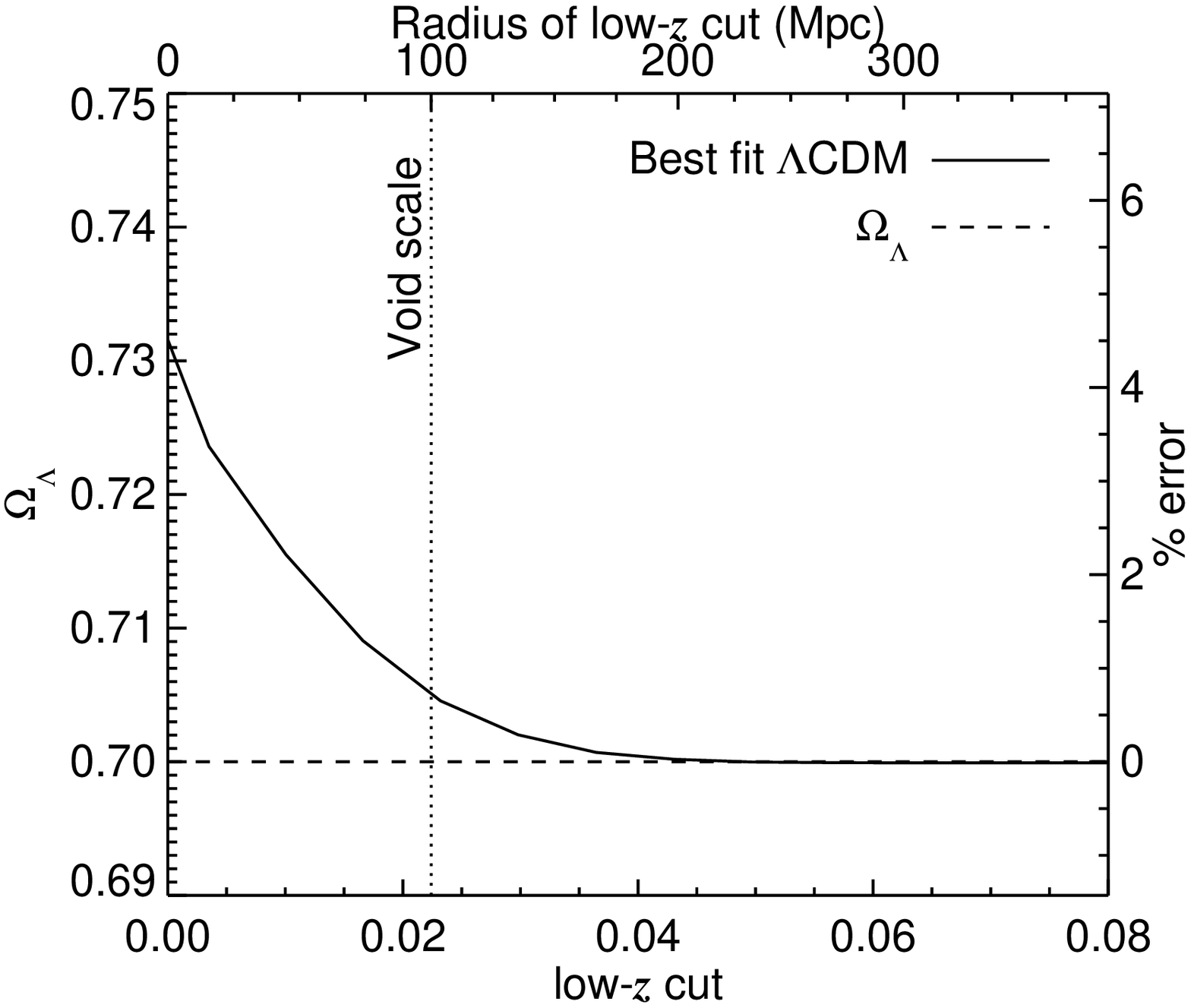}}
\end{center}
\caption{Same as Fig.~\ref{fig:BigVoid_z0_d03}, but for the realistic sized void considered in Sect.~\ref{sect:quantifying}.  The effect is smaller, but even for this size void the error is not negligible given the expected precision of upcoming surveys. }
\label{fig:RealLambda_z0_d03}
\end{figure}

We repeat this analysis for the Small-$\Lambda$ case. Figure~\ref{fig:RealLambda_OMz0_d03} shows that for this smaller under-density, the best fit mass density and cosmological constant density asymptote to the actual external densities as the low-z cutoff is increased. However, at the low redshift cutoff currently in use, $z=0.02$, the error in $\Omega_M$ is 0.06\%, and the error in $\OL$ is 0.99\%, the latter of which remains significant.

\begin{figure}
\epsscale{1.0}
\plotone{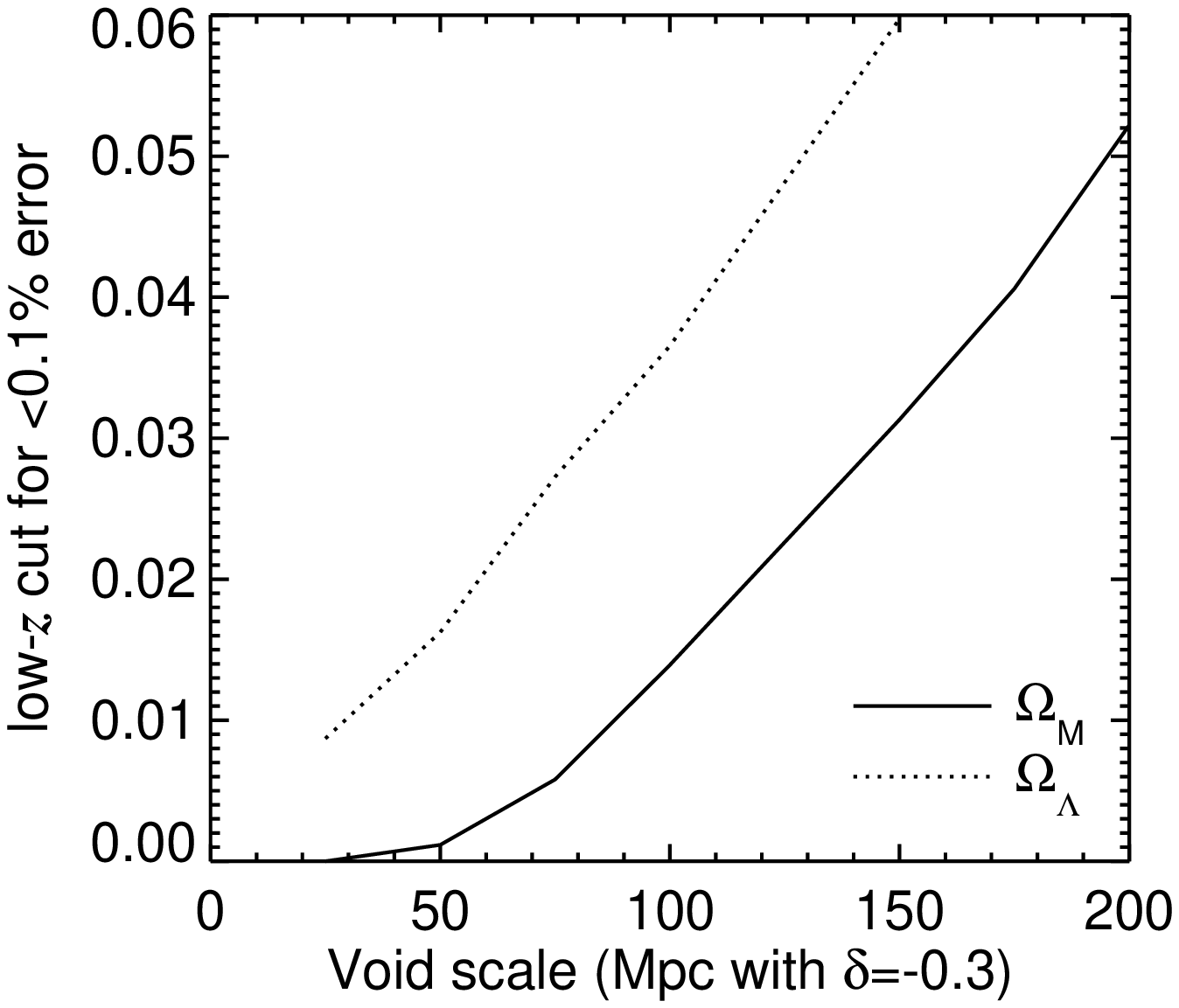}
\caption{This plot shows the low-$z$ cutoff needed to reduce the error in $\OM$ and $\OL$ to 0.1\% for various sized voids.  The more stringent constraint comes from $\OL$. }
\label{fig:Various_Voids_size}
\end{figure}

\begin{figure}
\epsscale{1.0}
\plotone{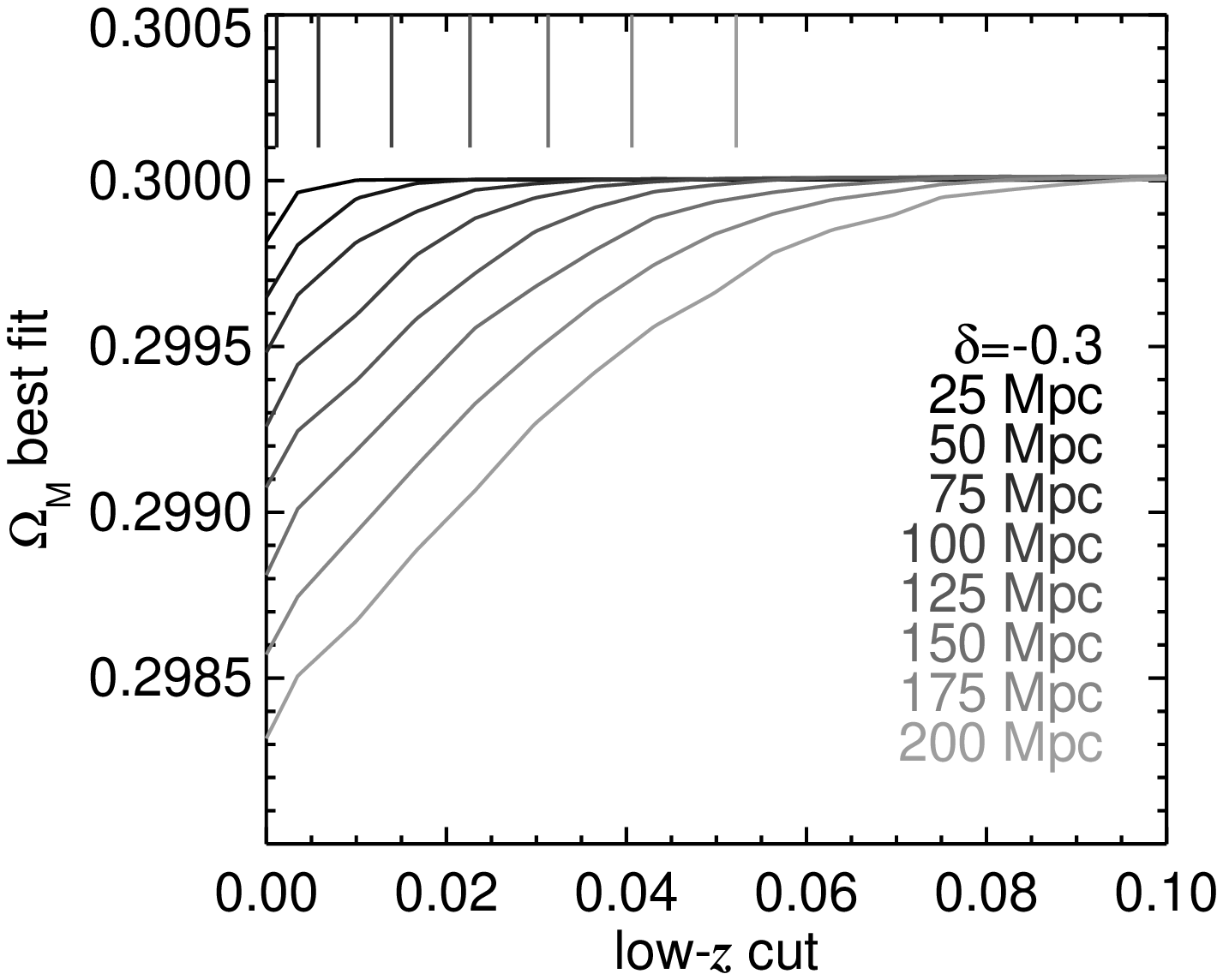}
\plotone{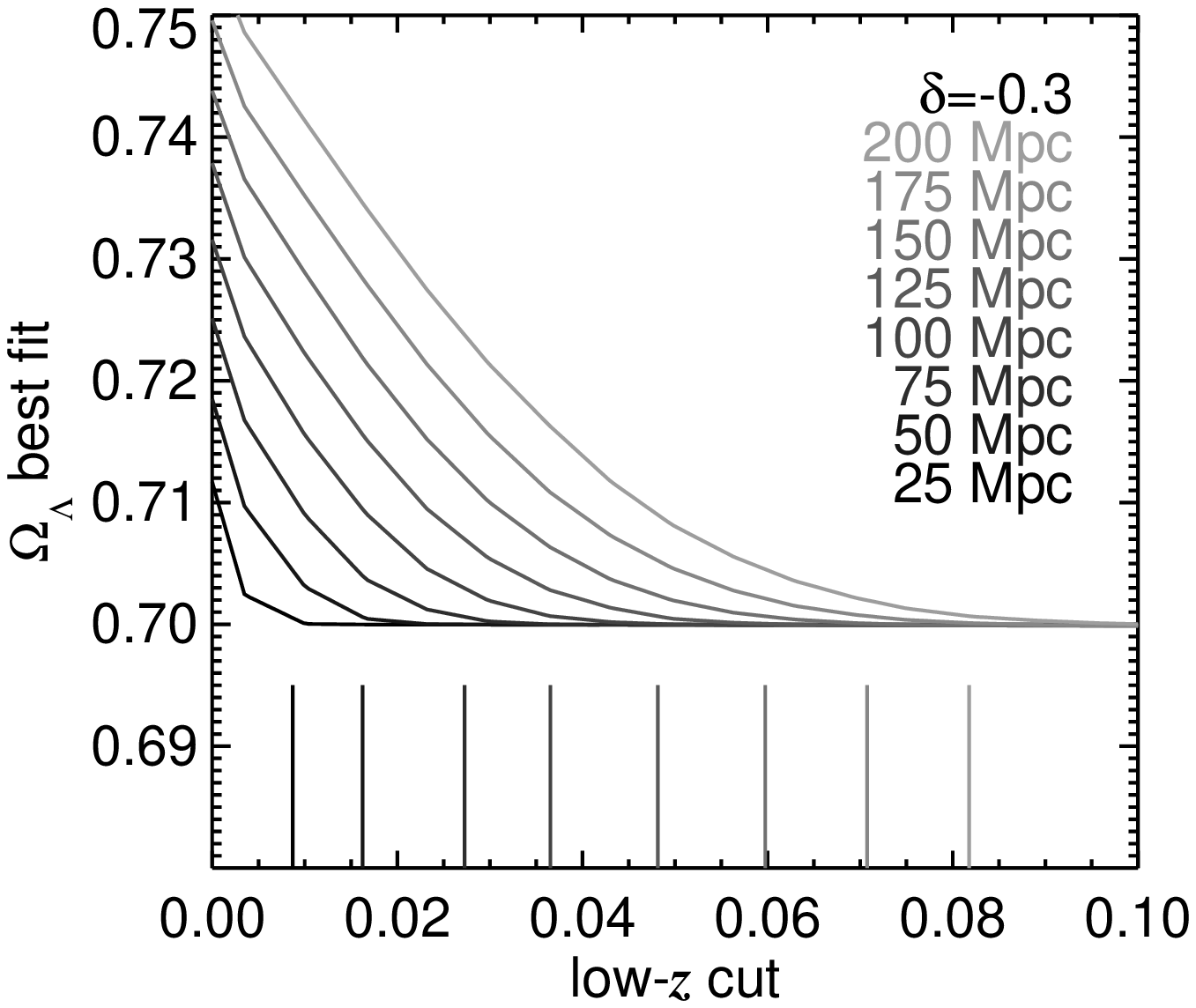}
\caption{How the best fit values of $\OM$ and $\OL$ change with low redshift cutoff for a range of void sizes.  As the void size increases the low-redshift cutoff needed to reduce the systematic error below a certain threshold increases.  The vertical lines represent the cutoffs needed to reduce the systematic error below 0.1\% in each case.  Void sizes tested range from 25 to 200 Mpc, and in each case the density contrast is taken to be $\delta=-0.3$.  These are considered amongst the plausible range of density fluctuations for our universe.}
\label{fig:Various_Voids_omol}
\end{figure}

An error of 0.06\% in $\OM$ is not likely to be a significant fraction of the observational error in the foreseeable future, but the error of 0.99\% in $\OL$ is already comparable to the observational error. This suggests that a higher low-$z$ cutoff may be advisable. Figure~\ref{fig:Various_Voids_size} shows the low-$z$ cutoff needed to reduce the errors in $\OM$ and $\OL$ to 0.1\% of the exterior value. Assuming a void size of 70\hinvMpc\ ($\sim$100Mpc), a low-$z$ cutoff of about 0.035 Mpc is needed to reduce the error in $\OL$ to 0.1\%. Of course this is very dependent on the size of void we chose to test.  We show in Fig.~\ref{fig:Various_Voids_omol} how the best fit $\OM$ and $\OL$ values change with the low-$z$ cutoff for a range of void scales.

\begin{figure}
\epsscale{1.0}
\plotone{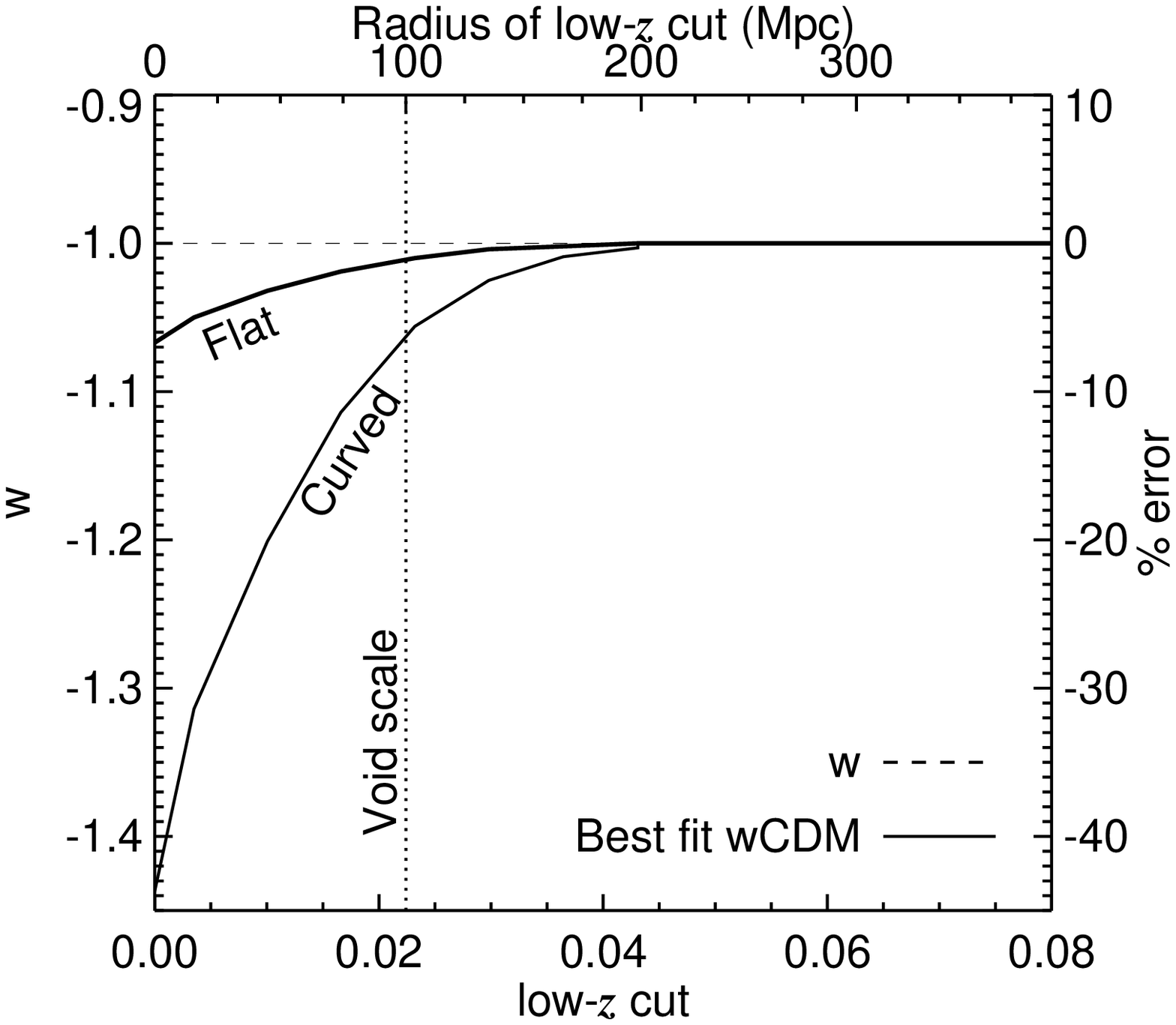}
\caption{As with $\OM$ and $\OL$, the under-density leads to an error in the best fit value of the parameter $w$, which is lessened as $z_0$ increases.  When the model is constrained to be flat the error in the best fit $w$ is smaller, primarily because the uncertainty in $w$ becomes larger.  Given the increased uncertainty in the curved case, the error is no more significant than the error when the model is constrained to be flat.} 
\label{fig:VariationOfBestFitWwreal_lambda}
\label{fig:VariationOfBestFitWwFlatreal_lambda}  
\end{figure}

Finally, Figure~\ref{fig:VariationOfBestFitWwreal_lambda} shows the best fit value of $w$ as a function of $z_0$. As was the case for $\OM$ and $\OL$, the best fit value is different to the true value, but converges to the true value as we increase $z_0$.  When all three of $\OM$, $\OL$, and $w$ are allowed to vary freely the error in $w$ can be reduced from 44\%, with no low-$z$ cutoff, to 1.2\% with a low-$z$ cutoff of 0.035.  At the currently preferred low-$z$ cutoff of 0.02 there remains an error of 8.5\% (again this is dependent on the void size chosen).   When we impose flatness by setting $\OM+\OL=1.0$ we effectively reduce the number of free parameters by one and as shown previously, the error even with $z_0=0$ is much reduced at 6.7\%.  When we also add a low-$z$ cutoff of 0.02 this is reduced to a 1.4\% error and a low-$z$ cutoff of 0.035 reduces the error to just 0.2\%.

\section{Discussion}\label{sect:discussion}

\subsection{Likelihood of Drawing Incorrect Conclusions}\label{sect:IncorrectConclusions}

We have seen throughout that the each time we allow a new parameter to vary, a better fit is found, but that the best fit parameters are altered such that they do not give an accurate representation of the cosmological parameters outside the under-dense region. For example, in the matter only LTB model, allowing $\OL$ to vary improves the fit but results in an $\OL{\neq}0$, despite the LTB model having $\OL=0$.

It is important to consider whether a cosmologist making such fits would consider an improvement in agreement with the data upon varying a new parameter, to be strong evidence for a value of that parameter different to that when it is not allowed to vary.
Fitting an additional parameter, as opposed to setting it to a fixed value, will always improve the goodness of fit of a test model, regardless of whether this parameter is physically motivated.  For example, a higher order polynomial will always give an equal or better fit than a linear function to any data set, even when the extra parameters are spurious. 

Clearly it does not give any extra insight to include extra parameters without any physical meaning. 
So to determine whether a parameter, or a particular value of a parameter does indeed well reflect the system under consideration, one can look at the improvement in the goodness of fit (GoF) resulting from varying the parameter. Here we use the definition of GoF as $\chi^2/${\em dof}, where {\em dof} is the number of degrees of freedom (number of data points minus number of parameters). If the goodness of fit improves by a large amount upon allowing a parameter to differ from a certain value, then this is taken as strong evidence that the underlying system is described by a parameter with value different to the previously set value. However, if the goodness of fit improves only a little, then it is likely that the improvement is simple due to the extra freedom, granted by the variation of the parameter, for a model to fit the data set and its associated errors.  

This is often quantified in ``Information Criteria" such as the Akaike information criterion, $AIC=-2\ln {\cal L} +2k$, where ${\cal L}$ represents the likelihood, $k$ represents the number of parameters, and for gaussian errors $-2\ln{\cal L} = \chi^2$ \citep{akaike74}.  A $\Delta AIC > 6$ between two models is considered significant evidence that any extra parameters in the model with the lower $AIC$ are well justified by the improvement in the fit.  In other words, it is unlikely that such a large improvement to the fit would occur unless the extra parameter described a genuine feature of the data. For a discussion of information criteria in a cosmological context  and the more rigorous Bayesian evidence see \citet{liddle04,liddle06}.

To yield goodness of fit calculations using our simulated data which would match those made by a cosmologist in an LTB universe, we fist add a normally distributed random error with standard deviation 0.2 mag to the distance modulus at each redshift, this replicates the observational error a measurement would incur. As before we then assign an uncertainty to this measurement of distance modulus of 0.2. With this simulated data we can then calculate the best fit cosmological parameters and the associated goodness of fit. To remove the dependence of the value of goodness of fit and best fit cosmological parameters on the particular set of random errors generated, we repeat this process 25 times with different sets of normally distributed random errors to get an average goodness of fit.

We first consider the model in section~\ref{sect:quantifying}, a small under-density embedded in a universe with both matter and cosmological constant. The best fit \lcdm to this model is $(\OM,\OL)=(0.305,0.72)$, with $\chi^2/${\em dof}=1.00. 
We note that the best fit matter and cosmological constant densities are slightly different to those stated in section~\ref{sect:quantifying} since here we have added many different sets of observational errors and taken an average, in section~\ref{sect:quantifying} we did not add observational errors, nor perform multiple runs.
 If we fit a set of models with only matter, $\OM=0$, we find that the best fit CDM is $(\OM,\OL)=(0.214,0.0)$, with $\chi^2/${\em dof}=1.07. Here we see that allowing $\OL$ to vary from 0, the goodness of fit is much improved ($\Delta$AIC=19), and the cosmologist would conclude that there is strong evidence that the universe has a cosmological constant with $\OL{\neq}0$, which is indeed the case. This shows that a model with $\OL=0$ is unable to give a good fit to the LTB model with cosmological constant, and allowing $\OL$ to vary is both required in order to obtain a good fit and does shed light on the nature of the actual universe i.e. predicts a non-zero cosmological constant. 

Now we give a counter example, where an improved fit should not be considered significant. In section~\ref{sect:largeVoid}, we obtained the best fit \lcdm model to a matter-only LTB model. With simulated errors, the average best fit parameters are $(\OM,\OL)=(0.85,0.41)$, where we have allowed $\OL$ to vary from 0, despite the LTB model having $\OL=0$. For this \lcdm, the $\chi^2/${\em dof} is 0.994 (AIC$=301.2$). Fitting a matter-only model (by setting $\OL=0$), we obtain best fit parameters of $(\OM,\OL)=(0.87,0.00)$, and a $\chi^2$/dof of 1.009 (AIC$=304.7$).  As before, there is a clear improvement in the goodness of fit by allowing $\OL$ to vary from 0, despite the fact that the model which generated the data had $\OL=0$. This is a case in which, solely due to the presence of the local under-density, the astronomer may incorrectly conclude that there is evidence for a cosmological constant with density $\OL{\neq}0$.  However, the $\Delta$AIC between the two models is only 3.5, which should only be considered weak evidence in support of the better fit.  In this case the AIC indicates that there is a non-negligible chance that the extra parameter is spurious.  

Finally, we consider whether the presence of a local under-density would induce an astronomer to incorrectly conclude that the equation of state of dark energy was different to $w=-1$. Using the Small${\Lambda}$ model as the data set, in section~\ref{sect:quantifying} we fitted a model where $w$ is fixed as $-1$. Doing so again including observational errors we obtain a best fit universe of $(\OM,\OL,w)=(0.305,0.72,-1)$ and a ($\chi^2$, $\chi^2/${\em dof}) of (299.0, 1.000). Allowing $w$ to vary away from -1, the best fit model, as described in section~\ref{sect:EoS}, is now one with parameters $(\OM,\OL,w)=(0.246,0.75,-1.65)$ and the $\chi^2$ is reduced to 298.2, whilst the $\chi^2/${\em dof} basically stays constant (it worsens very slightly to 1.001).  Despite the drastic deviation from the $w=-1$ model, the best fit is only a small improvement and $\Delta$AIC$= -3.2$.  Thus we would conclude that there is no evidence for an equation of state with $w\ne -1$, but rather evidence that the extra parameter gave too much freedom to the model leaving it poorly constrained.  

Thus we conclude that a void gives a high-risk of fooling us into believing an incorrect $\OL$ but gives a low-risk of fooling us into believing $w\ne-1$.  However, once we constrain to flatness we note that Flat-$w$CDM and $\Lambda$CDM have the same number of parameters and are equally good fits to the data, so we have no statistical reason to prefer one over the other.

Table~\ref{table:AddingLambda} summarizes the results presented in this section. They show the improvement in goodness of fit for each extra parameter added, which as discussed gives an indication of our susceptibility to be deceived by the void.

\subsection{Intuitive understanding of the void effect.}\label{sect:intuitive}
The under-dense region has two effects on light passing through it. Firstly, the lower density gives rise to a greater rate of expansion, $H_{\rm in}>H_{\rm out}$, because of less gravitating mass. Thus light traveling through this region is cosmologically redshifted more than it would have been had it traveled through a medium of $\OM=\OMout$. In addition, the under-dense region is at a higher gravitational potential than the external medium, so it causes an additional gravitational redshift to the light immediately before we observe it. Both these effects cause incoming light to be more redshifted than had it simply traveled through a homogeneous medium of density $\OM=\OMout$. Hence the \lcdm model we fit to this data will be one that redshifts light more than one with $\OM=\OMout$.  This behavior is characteristic of a model with $\OM<\OMout$ and $\OL>0$.  Thus by not accounting for a Hubble-bubble we mistakenly deduce a lower-than-actual matter density together with the existence of a cosmological constant.

\begin{figure}
\epsscale{1.0}
\plotone{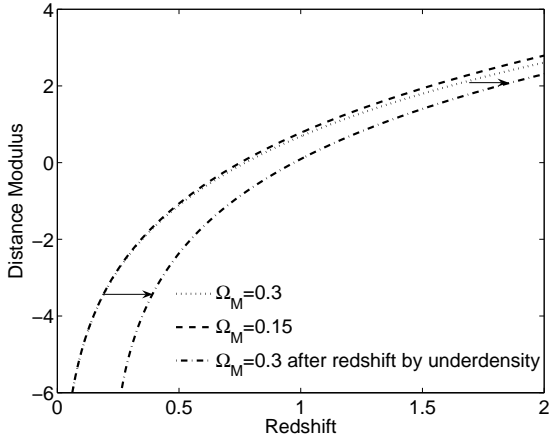}
\caption{When there is an under-density in the local universe all incoming data is redshifted.  This is why removing the low-$z$ data that come from inside the void does not allow us to completely recover the correct model.  By effectively shifting all the points in the magnitude redshift diagram to higher redshift the void has changed the slope of the relation, which is the feature we're fitting.  Therefore as we progressively remove data points from low z, the estimation of the external matter density improves, but does not converge to the true value.}
\label{fig:whichFitsBetterNow}
\end{figure}

In Fig.~\ref{fig:whichFitsBetterNow} we show diagrammatically how a low-$z$ cutoff is inherently incapable of removing the effect entirely.
The extra redshift due to the void will be the same for all light sources situated outside the void. The effect is to shift the magnitude-redshift relation along in the positive $z$ direction. In this figure we see that at each value of $z$ the magnitude-redshift relation distorted in this way has a gradient more similar to that of a \lcdm model with a lower $\OM$.

This graphical interpretation of the effect is equivalent to realizing that even high redshift sources are affected by the under-dense region. This is apparent from the outset, since light from far away still has to pass through the under-dense region in order to reach us and be observed, and hence will incur a greater than expected redshift, if we do not take the local inhomogeneity into account. Thus, it is evident that the current practice of neglecting sources at low redshifts in the hope of removing the effect of the void is optimistic, but as we have shown quantitatively, this approach does have the desired effect provided that the under-density is of the scale of observed inhomogeneity in our universe. For larger voids, removing supernova observations from low redshifts will not remove the effect of the void, regardless of how much data we neglect, as demonstrated in sections~\ref{sect:largeVoid} and~\ref{sect:largeLambda}.

Although we have only simulated under-densities in this paper the same principle applies to over-densities. 

We have also tested whether the asymptotic value of $\OM$ relates to any average matter density in the volume we can define, including the lower density of the void.  The answer is no. 

\section{Conclusion}\label{sect:conclusion}

We have shown that a local density fluctuation can leave a significant systematic error on our cosmological inferences from distance measures.  This applies not only to supernovae, as discussed here, but also to measures of angular diameter distance such as baryon acoustic oscillations and the cosmic microwave background.  The low-$z$ cutoff imposed on supernova data is absolutely crucial to avoid significant systematic errors, far above the statistical uncertainty in current observations. This is particularly true for the cosmological constant, whose form can mimic that of a local void. We have shown that for an under-density with a scale and depth typical of voids in the current standard model and the low-$z$ cutoff currently in use, z=0.02, a systematic error of 0.99\% remains in the deduced value of $\OL$ and hence a slightly higher low-$z$ cutoff of 0.035 is advisable, at which the error drops to a negligible 0.1\%. The equation of state of dark energy is less susceptible to bias induced by a void than a cosmological constant, because it does not easily mimic the void effect, but a local under-density can still induce wild discrepancies in the best-fit values.  

We have also shown that introducing a low-$z$ cutoff can reduce {\em but not remove} the systematic error on the inferred cosmological parameters.   Thus a void imprints an irreducible error that can only be removed by fitting the extra parameters needed to allow for a local density fluctuation.  The magnitude of this bias is small compared to the current observational uncertainty on $\OM$, $\OL$, and $w$.  Nevertheless, as our cosmological inferences become more precise, this bias will no longer be negligible and will need to be accounted for.


\acknowledgments

\newpage
\bibliographystyle{hapj}
\bibliography{bib_sinc09}

\begin{thebibliography}{31}
\expandafter\ifx\csname natexlab\endcsname\relax\def\natexlab#1{#1}\fi

\bibitem[{{Akaike}(1974)}]{akaike74}
{Akaike}, H. 1974, IEEE Transactions on Automatic Control, 19, 716

\bibitem[{{Alexander} {et~al.}(2009){Alexander}, {Biswas}, {Notari}, \&
  {Vaid}}]{alexander09}
{Alexander}, S., {Biswas}, T., {Notari}, A., \& {Vaid}, D. 2009, Journal of
  Cosmology and Astro-Particle Physics, 9, 25, 0712.0370

\bibitem[{{Alnes} \& {Amarzguioui}(2006)}]{alnes06b}
{Alnes}, H., \& {Amarzguioui}, M. 2006, \prd, 74, 103520,
  arXiv:astro-ph/0607334

\bibitem[{{Alnes} {et~al.}(2006){Alnes}, {Amarzguioui}, \&
  {Gr{\o}n}}]{alnes06a}
{Alnes}, H., {Amarzguioui}, M., \& {Gr{\o}n}, {\O}. 2006, \prd, 73, 083519,
  arXiv:astro-ph/0512006

\bibitem[{{Astier} {et~al.}(2006){Astier}, {Guy}, {Regnault}, {Pain},
  {Aubourg}, {Balam}, {Basa}, {Carlberg}, {Fabbro}, {Fouchez}, {Hook},
  {Howell}, {Lafoux}, {Neill}, {Palanque-Delabrouille}, {Perrett}, {Pritchet},
  {Rich}, {Sullivan}, {Taillet}, {Aldering}, {Antilogus}, {Arsenijevic},
  {Balland}, {Baumont}, {Bronder}, {Courtois}, {Ellis}, {Filiol}, {Gon{\c
  c}alves}, {Goobar}, {Guide}, {Hardin}, {Lusset}, {Lidman}, {McMahon},
  {Mouchet}, {Mourao}, {Perlmutter}, {Ripoche}, {Tao}, \& {Walton}}]{astier06}
{Astier}, P. {et~al.} 2006, \aap, 447, 31, astro-ph/0510447

\bibitem[{{Clifton} {et~al.}(2008){Clifton}, {Ferreira}, \& {Land}}]{clifton08}
{Clifton}, T., {Ferreira}, P.~G., \& {Land}, K. 2008, Physical Review Letters,
  101, 131302, 0807.1443

\bibitem[{{Conley} {et~al.}(2007){Conley}, {Carlberg}, {Guy}, {Howell}, {Jha},
  {Riess}, \& {Sullivan}}]{conley07}
{Conley}, A., {Carlberg}, R.~G., {Guy}, J., {Howell}, D.~A., {Jha}, S.,
  {Riess}, A.~G., \& {Sullivan}, M. 2007, \apjl, 664, L13, 0705.0367

\bibitem[{{Enqvist}(2008)}]{enqvist08}
{Enqvist}, K. 2008, General Relativity and Gravitation, 40, 451, 0709.2044

\bibitem[{{Furlanetto} \& {Piran}(2006)}]{furlanetto06}
{Furlanetto}, S.~R., \& {Piran}, T. 2006, \mnras, 366, 467,
  arXiv:astro-ph/0509148

\bibitem[{{Garc{\'{\i}}a-Bellido} \&
  {Haugb{\o}lle}(2008{\natexlab{a}})}]{garcia-bellido08}
{Garc{\'{\i}}a-Bellido}, J., \& {Haugb{\o}lle}, T. 2008{\natexlab{a}}, Journal
  of Cosmology and Astro-Particle Physics, 4, 3, 0802.1523

\bibitem[{{Garc{\'{\i}}a-Bellido} \&
  {Haugb{\o}lle}(2008{\natexlab{b}})}]{garcia-bellido08-kz}
------. 2008{\natexlab{b}}, Journal of Cosmology and Astro-Particle Physics, 9,
  16, 0807.1326

\bibitem[{{Garc{\'{\i}}a-Bellido} \& {Haugb{\o}lle}(2009)}]{garcia-bellido09}
------. 2009, Journal of Cosmology and Astro-Particle Physics, 9, 28, 0810.4939

\bibitem[{{Geller} \& {Huchra}(1989)}]{geller89}
{Geller}, M.~J., \& {Huchra}, J.~P. 1989, Science, 246, 897

\bibitem[{{Geller} {et~al.}(1997){Geller}, {Kurtz}, {Wegner}, {Thorstensen},
  {Fabricant}, {Marzke}, {Huchra}, {Schild}, \& {Falco}}]{geller97}
{Geller}, M.~J. {et~al.} 1997, \aj, 114, 2205, arXiv:astro-ph/9710109

\bibitem[{{Giannantonio} {et~al.}(2008){Giannantonio}, {Scranton},
  {Crittenden}, {Nichol}, {Boughn}, {Myers}, \& {Richards}}]{giannantonio08}
{Giannantonio}, T., {Scranton}, R., {Crittenden}, R.~G., {Nichol}, R.~C.,
  {Boughn}, S.~P., {Myers}, A.~D., \& {Richards}, G.~T. 2008, \prd, 77, 123520,
  0801.4380

\bibitem[{{Giovanelli} {et~al.}(1999){Giovanelli}, {Dale}, {Haynes}, {Hardy},
  \& {Campusano}}]{giovanelli99}
{Giovanelli}, R., {Dale}, D.~A., {Haynes}, M.~P., {Hardy}, E., \& {Campusano},
  L.~E. 1999, \apj, 525, 25, arXiv:astro-ph/9906362

\bibitem[{{Haugb{\o}lle} {et~al.}(2007){Haugb{\o}lle}, {Hannestad}, {Thomsen},
  {Fynbo}, {Sollerman}, \& {Jha}}]{haugbolle07}
{Haugb{\o}lle}, T., {Hannestad}, S., {Thomsen}, B., {Fynbo}, J., {Sollerman},
  J., \& {Jha}, S. 2007, \apj, 661, 650, arXiv:astro-ph/0612137

\bibitem[{{Hoyle} \& {Vogeley}(2004)}]{hoyle04}
{Hoyle}, F., \& {Vogeley}, M.~S. 2004, \apj, 607, 751, arXiv:astro-ph/0312533

\bibitem[{{Hudson} {et~al.}(2004){Hudson}, {Smith}, {Lucey}, \&
  {Branchini}}]{hudson04}
{Hudson}, M.~J., {Smith}, R.~J., {Lucey}, J.~R., \& {Branchini}, E. 2004,
  \mnras, 352, 61, arXiv:astro-ph/0404386

\bibitem[{{Jha} {et~al.}(2007){Jha}, {Riess}, \& {Kirshner}}]{jha07}
{Jha}, S., {Riess}, A.~G., \& {Kirshner}, R.~P. 2007, \apj, 659, 122,
  arXiv:astro-ph/0612666

\bibitem[{{Kashlinsky} {et~al.}(2008){Kashlinsky}, {Atrio-Barandela},
  {Kocevski}, \& {Ebeling}}]{kashlinsky08}
{Kashlinsky}, A., {Atrio-Barandela}, F., {Kocevski}, D., \& {Ebeling}, H. 2008,
  \apjl, 686, L49, 0809.3734

\bibitem[{{Kessler} {et~al.}(2009){Kessler}, {Becker}, {Cinabro}, {Vanderplas},
  {Frieman}, {Marriner}, {Davis}, {Dilday}, {Holtzman}, {Jha}, {Lampeitl},
  {Sako}, {Smith}, {Zheng}, {Nichol}, {Bassett}, {Bender}, {Depoy}, {Doi},
  {Elson}, {Filippenko}, {Foley}, {Garnavich}, {Hopp}, {Ihara}, {Ketzeback},
  {Kollatschny}, {Konishi}, {Marshall}, {McMillan}, {Miknaitis}, {Morokuma},
  {M{\"o}rtsell}, {Pan}, {Prieto}, {Richmond}, {Riess}, {Romani}, {Schneider},
  {Sollerman}, {Takanashi}, {Tokita}, {van der Heyden}, {Wheeler}, {Yasuda}, \&
  {York}}]{kessler09}
{Kessler}, R. {et~al.} 2009, \apjs, 185, 32, 0908.4274

\bibitem[{{Kim} {et~al.}(2004){Kim}, {Linder}, {Miquel}, \& {Mostek}}]{kim04}
{Kim}, A.~G., {Linder}, E.~V., {Miquel}, R., \& {Mostek}, N. 2004, \mnras, 347,
  909, arXiv:astro-ph/0304509

\bibitem[{{Komatsu} {et~al.}(2008){Komatsu}, {Dunkley}, {Nolta}, {Bennett},
  {Gold}, {Hinshaw}, {Jarosik}, {Larson}, {Limon}, {Page}, {Spergel},
  {Halpern}, {Hill}, {Kogut}, {Meyer}, {Tucker}, {Weiland}, {Wollack}, \&
  {Wright}}]{komatsu08}
{Komatsu}, E. {et~al.} 2008, ArXiv e-prints, 803, 0803.0547

\bibitem[{{Kowalski} {et~al.}(2008){Kowalski}, {Rubin}, {Aldering},
  {Agostinho}, {Amadon}, {Amanullah}, {Balland}, {Barbary}, {Blanc}, {Challis},
  {Conley}, {Connolly}, {Covarrubias}, {Dawson}, {Deustua}, {Ellis}, {Fabbro},
  {Fadeyev}, {Fan}, {Farris}, {Folatelli}, {Frye}, {Garavini}, {Gates},
  {Germany}, {Goldhaber}, {Goldman}, {Goobar}, {Groom}, {Haissinski}, {Hardin},
  {Hook}, {Kent}, {Kim}, {Knop}, {Lidman}, {Linder}, {Mendez}, {Meyers},
  {Miller}, {Moniez}, {Mour{\~a}o}, {Newberg}, {Nobili}, {Nugent}, {Pain},
  {Perdereau}, {Perlmutter}, {Phillips}, {Prasad}, {Quimby}, {Regnault},
  {Rich}, {Rubenstein}, {Ruiz-Lapuente}, {Santos}, {Schaefer}, {Schommer},
  {Smith}, {Soderberg}, {Spadafora}, {Strolger}, {Strovink}, {Suntzeff},
  {Suzuki}, {Thomas}, {Walton}, {Wang}, {Wood-Vasey}, \& {Yun}}]{kowalski08}
{Kowalski}, M. {et~al.} 2008, \apj, 686, 749, 0804.4142

\bibitem[{{Liddle}(2004)}]{liddle04}
{Liddle}, A.~R. 2004, \mnras, 351, L49, astro-ph/0401198

\bibitem[{{Liddle} {et~al.}(2006){Liddle}, {Mukherjee}, {Parkinson}, \&
  {Wang}}]{liddle06}
{Liddle}, A.~R., {Mukherjee}, P., {Parkinson}, D., \& {Wang}, Y. 2006, ArXiv
  Astrophysics e-prints, astro-ph/0610126

\bibitem[{{Sollerman} {et~al.}(2009){Sollerman}, {M\"ortsell}, {Davis}, \&
  {SDSS}}]{sollerman09}
{Sollerman}, J., {M\"ortsell}, E., {Davis}, T.~M., \& {SDSS}. 2009, \apj

\bibitem[{{Wood-Vasey} {et~al.}(2007){Wood-Vasey}, {Miknaitis}, {Stubbs},
  {Jha}, {Riess}, {Garnavich}, {Kirshner}, {Aguilera}, {Becker}, {Blackman},
  {Blondin}, {Challis}, {Clocchiatti}, {Conley}, {Covarrubias}, {Davis},
  {Filippenko}, {Foley}, {Garg}, {Hicken}, {Krisciunas}, {Leibundgut}, {Li},
  {Matheson}, {Miceli}, {Narayan}, {Pignata}, {Prieto}, {Rest}, {Salvo},
  {Schmidt}, {Smith}, {Sollerman}, {Spyromilio}, {Tonry}, {Suntzeff}, \&
  {Zenteno}}]{WV07}
{Wood-Vasey}, W.~M. {et~al.} 2007, ArXiv Astrophysics e-prints,
  astro-ph/0701041

\bibitem[{{Zehavi} {et~al.}(1998){Zehavi}, {Riess}, {Kirshner}, \&
  {Dekel}}]{zehavi98}
{Zehavi}, I., {Riess}, A.~G., {Kirshner}, R.~P., \& {Dekel}, A. 1998, \apj,
  503, 483, astro-ph/9802252

\bibitem[{{Zibin} {et~al.}(2008){Zibin}, {Moss}, \& {Scott}}]{zibin08}
{Zibin}, J.~P., {Moss}, A., \& {Scott}, D. 2008, Physical Review Letters, 101,
  251303, 0809.3761

\end{thebibliography}


\clearpage

\end{document}